\documentclass[fleqn,usenatbib]{mnras}

\usepackage{newtxtext,newtxmath}

\usepackage[T1]{fontenc}

\DeclareRobustCommand{\VAN}[3]{#2}
\let\VANthebibliography\thebibliography
\def\thebibliography{\DeclareRobustCommand{\VAN}[3]{##3}\VANthebibliography}

\hypersetup{
    colorlinks=true,
    linkcolor=darkcolor,
    filecolor=darkcolor,
    urlcolor=darkcolor,
    citecolor=darkcolor
}
\usepackage{ulem}
\usepackage{caption}
\usepackage{subcaption}
\usepackage{newtxtext,newtxmath,tikz}
\usepackage{float}
\floatstyle{plaintop}
\restylefloat{table}
\usepackage[utf8]{inputenc}
\usepackage[T1]{fontenc}
\usepackage{ae,aecompl}
\usepackage{bm}
\usepackage{xcolor}
\usepackage{soul}
\usepackage{graphicx}
\usepackage{if then}
\usepackage{longtable}
\usetikzlibrary{shapes,arrows}
\usepackage{todonotes}
\usepackage{url}
\usepackage{mathtools}
\usepackage{xspace}

\definecolor{darkcolor}{rgb}{0.422,0.222,0.613} 
\definecolor{nicegreen}{HTML}{2CA02C}
\definecolor{darkgreen}{rgb}{0.0,0.5,0.0}

\newcommand{\sdssiii}{{SDSS-III/BOSS}}
\newcommand{\borg}{\textsc{borg}\xspace}

\newcommand{\fnl}{{f_\mathrm{nl}}}

\newcommand{\mvec}[1]{{\mathbf{#1}}}

\newcommand{\Mpch}{\ensuremath{h^{-1}\;\text{Mpc}}}

\title[Field-level inference of PNG]{Bayesian field-level inference of primordial non-Gaussianity using next-generation galaxy surveys}

\author[A. Andrews, J. Jasche, G. Lavaux, F. Schmidt]{
Adam Andrews$^{1}$\thanks{E-mail: adam.andrews@fysik.su.se},
Jens Jasche$^{1,2}$,
Guilhem Lavaux$^{2}$,
Fabian~Schmidt$^{3}$
\\
$^{1}$The Oskar Klein Centre, Department of Physics, Stockholm University, AlbaNova University Centre,
SE 106 91 Stockholm, Sweden\\
$^{2}$CNRS \& Sorbonne Université, Institut d'Astrophysique de Paris (IAP),  UMR 7095, 98 bis bd Arago, 75014 Paris, France \\
$^{3}$Max–Planck–Institut f\"ur Astrophysik, Karl–Schwarzschild–Stra\ss e 1, 85748 Garching, Germany
}

\date{Accepted XXX. Received YYY; in original form ZZZ}

\pubyear{2022}

\begin{document}
\label{firstpage}
\pagerange{\pageref{firstpage}--\pageref{lastpage}}
\maketitle

\begin{abstract}
Detecting and measuring a non-Gaussian signature of primordial origin in the density field is a major science goal of next-generation galaxy surveys. The signal will permit us to determine primordial physics processes and constrain models of cosmic inflation. While traditional approaches utilise a limited set of statistical summaries of the galaxy distribution to constrain primordial non-Gaussianity, we present a field-level approach by Bayesian forward-modelling the entire three-dimensional galaxy survey. Our method naturally and fully self-consistently exploits the entirety of the large-scale structure, e.g., higher-order statistics, peculiar velocity fields, and scale-dependent galaxy bias, to extract information on the local non-Gaussianity parameter, $\fnl$. We demonstrate the performance of our approach through various tests with mock galaxy data emulating relevant features of the \sdssiii{}-like survey, and additional tests with a \textit{Stage IV} mock data set. These tests reveal that the method infers unbiased values of $\fnl$ by accurately handling survey geometries, noise, and unknown galaxy biases. We demonstrate that our method can achieve constraints of $\sigma_{\fnl} \approx 8.78$ for \sdssiii{}-like data, an improvement of a factor $\sim 2.5$ over currently published constraints. Tests with next-generation mock data show that significant further improvements are feasible with sufficiently high resolution. Furthermore, the results demonstrate that our method can consistently marginalise all nuisance parameters of the data model. The method further provides an inference of the three-dimensional primordial density field, providing opportunities to explore additional signatures of primordial physics.
\end{abstract}


\section{Introduction}
\label{INTRO}

Deviations of primordial curvature fluctuations from a Gaussian random field, known as Primordial Non-Gaussianity (PNG), can constrain primordial universe physics and shed light on the number of fields driving inflation \citep{2003JHEP...05..013M,alvarez_testing_2014,2010AdAst2010E..72C,2010CQGra..27l4010K,2018arXiv181208197C,Senatore_2010, Barnaby_2012,Chen_2010}. The deviation from Gaussian initial conditions is commonly parameterised by the nonlinearity parameter, $f_{\mathrm{nl}}^{\mathrm{local}}$, denoted as $\fnl$ from now on. To date, the strongest constraint on $\fnl$ is based on observations of the Cosmic Microwave Background (CMB). CMB observations obtained by the Planck satellite yielded $\fnl = -0.9 \pm 5.1$ \citep[][]{planck_collaboration_planck_2019_IX}. While the CMB is a powerful cosmological probe, its information content at the largest scales has reached its cosmic variance limit, and constraints of order unity, $|\fnl| < 1$, are expected to be beyond its reach \citep[][]{Camera_2013,ballardini_constraining_2019,Moradinezhad_Dizgah_2019,Karagiannis_2020,meerburg_primordial_2019}.

For current state-of-the-art techniques which constrain PNG with the Large-Scale Structure (LSS), the most informative probe measures the two-point correlation function of galaxy populations. The reason for this is the scale-dependent bias, which yields the most information from the largest scales \citep{dalal_imprints_2008, Slosar_2008}. Previous works have been successful in constraining PNG in available galaxy redshift surveys \citep{Slosar_2008,Ross_2012,leistedt_constraints_2014,2018MNRAS.474.2853U,10.1093/mnras/stu590,2019MNRAS.485.4160M,castorina_redshift-weighted_2019,mueller2021clustering,cabass_constraints_2022,damico_limits_2022}; the tightest large-scale structure measurements comes from the SDSS-IV (DR 16, quasar sample) data, reaching a constraining power of $|\fnl| < 21$ \citep{mueller2021clustering}. 

Next-generation galaxy redshift surveys (\textit{Vera C. Rubin Observatory} \citep[]{lsst_science_collaboration_lsst_2009}, \textit{Euclid} \citep[]{2018LRR....21....2A}, and \textit{SPHEREx} \citep{dore_cosmology_2014}) will study the Universe at new levels of precision. These missions will be dominated by their systematic effects, instead of by their statistical power \citep{Graham_2018}. Thus, a high degree of mitigation and modelling of the survey effects, instrumentation noise, and astrophysical contamination are required. If unaccounted for, these effects can bias the results on the largest scales,  \citep{huterer2013,leistedt_constraints_2014,ho2015,jasche2017,2019A&A...625A..64J}.

A promising approach to analyse large-scale structure data is the field-level inference method \citep{jasche_bayesian_2010,Wang:2014hia,2017JCAP...12..009S,Lavaux2019,schmittfull2019,2020A&A...642A.139P,Schmidt:2020viy,2020JCAP...12..011N}. This method has the goal of going beyond individual correlation functions and extract the maximum amount of information to solve cosmological problems, e.g., constraining cosmological parameters \citep[]{ramanah_cosmological_2019,Leclercq_2021}. Since the full cosmological map is included into the analysis, a field-level inference can incorporate other informative probes into the analysis, e.g., peculiar velocity fields and higher-order statistics in the cosmic density field \citep{biagetti_hunt_2019,baumann2021power}.

\begin{figure}
	\center
	\includegraphics[width=1.0\columnwidth]{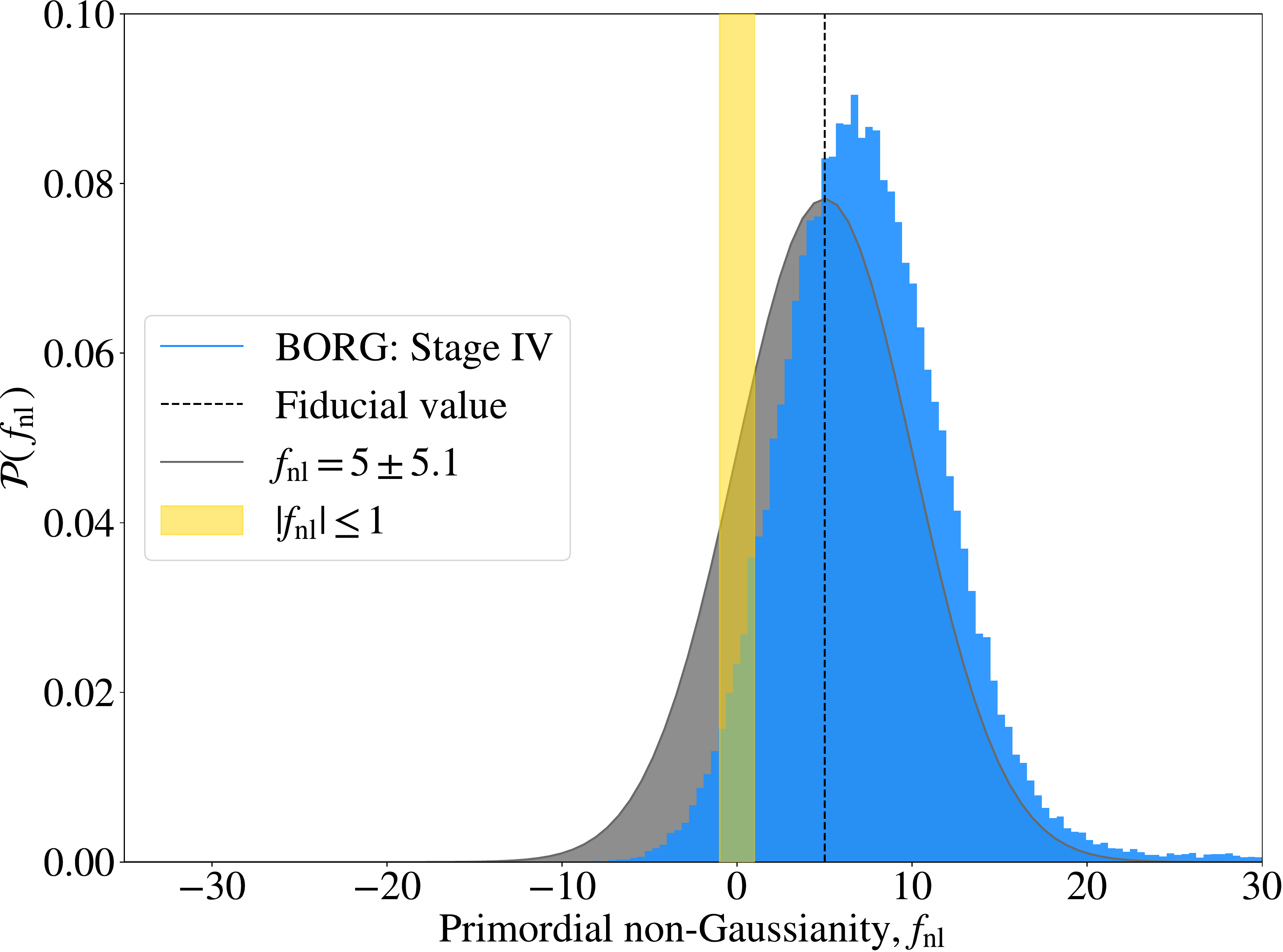}
	\caption{Conditional posterior distribution for the local type primordial non-Gaussianity parameter $\fnl$ obtained from a \borg{} application to a \textit{Stage IV} mock galaxy survey as demonstrated in this paper. This posterior incorporates all signatures of primordial non-Gaussianity in the cosmic large-scale structure, for the scales included in this project. Furthermore, our method correctly handles systematic errors and survey effects, as well as marginalising out bias parameters.The yellow band indicates the region $|\fnl| \leq 1$, the target region of next-generation surveys. To compare with the current best constraints, we plot a Gaussian centered on the fiducial value of $\fnl$, with a width corresponding to the Planck18 constraints \protect\citep{planck_collaboration_planck_2019_X}.}
	\label{fig:fnl_histogram}
\end{figure}

In this paper, we present for the first time a proof-of-concept to infer PNG in galaxy redshift surveys, with the aim of addressing current open issues. Our method is based on the physical forward modelling of the three-dimensional galaxy distribution in a Bayesian hierarchical framework. This implies that the algorithm performs a field-level inference of the full cosmic density field \citep{jasche_bayesian_2013}. In this way, our method self-consistently and naturally accounts for the current phenomenology of PNG when analysing galaxy data, and uses all of the information available in the cosmic density field when exploring the solution space \citep{jasche_bayesian_2013}. The physics-informed inference algorithm provides a path to account for various uncertainties and observational and systematic effects which otherwise can bias the cosmological conclusions drawn from data. These effects include survey geometry, selection effects, instrumentation noise, galaxy biases and foreground contamination of data \citep{jasche2017, 2019A&A...624A.115P}. Furthermore, the algorithm naturally accounts for higher-order statistics and filamentary cosmic structure associated with nonlinear structure formation, and can incorporate any probe associated with PNG. As a result, our method constrains $\fnl$ beyond the current state-of-the-art methods, and can place constraints as low as $|\fnl| < 5.70$, for mock data emulating features of next-generation galaxy redshift surveys (see Fig. \ref{fig:fnl_histogram}), while marginalising out systematic effects.

A major advantage of the forward modelling approach is that the full 3D density field is a part of the analysis \citep[]{jasche_bayesian_2013,baumann2021power}. Hence, \borg{} naturally and fully self-consistently accounts for a variety of probes when constraining $\fnl$:
\begin{enumerate}
    \item higher-order statistics of primordial origin in the dark matter density field \citep[]{Baldauf_2011,Tasinato_2014}
    \item mass distributions of the large-scale structure (for example, statistical moments such as skewness and kurtosis of the galaxies) \citep[][]{chodorowski_kurtosis_1996,durrer_skewness_2000,dalal_imprints_2008,yokoyama_modification_2011,friedrich_primordial_2019}
    \item peculiar velocity fields \citep[][]{schmidt_large-scale_2010,catelan_velocity_1995,ma_independent_2013,lam_pairwise_2011}
    \item scale-dependent galaxy bias \citep[]{castorina_redshift-weighted_2019,de_putter_designing_2017}
\end{enumerate}
For more details on the listed probes, the interested reader is referred to reviews found in the literature \citep[]{biagetti_hunt_2019}.
Another advantage is the fact that our method allows us to simultaneously analyse effects that are degenerate in other analysis methods. In our physics model (see Section \ref{Method}), the input field is iteratively built upon to create a model prediction. Thus, different effects enter the field at different locations in the pipeline, meaning that these effects can be disentangled. For example, while both structure formation and the perturbation of the primordial gravitational potential both affect the bispectrum (or three-point correlation function) of the density field, \borg{} is able to jointly explore these two effects in its framework, because \borg{} explicitly accounts for the nonlinear gravitational evolution.

The paper is structured as follows. We describe the Bayesian Origin Reconstruction from Galaxies (\borg{}) algorithm together with the developments made to the physical model to constrain $\fnl$ in Section \ref{Method}. In Section \ref{data_section}, we describe the artificial mock data set generation, detailing both the \sdssiii{}-like mock data set and the \textit{Stage IV} mock data set. Results are outlined in Section \ref{results}, where we evaluate the performance of the PNG-inference framework within \borg{}. We provide concluding thoughts and a summary in Section \ref{conclusions}.

\begin{figure*}
	\centering
    \includegraphics[width=2.0\columnwidth]{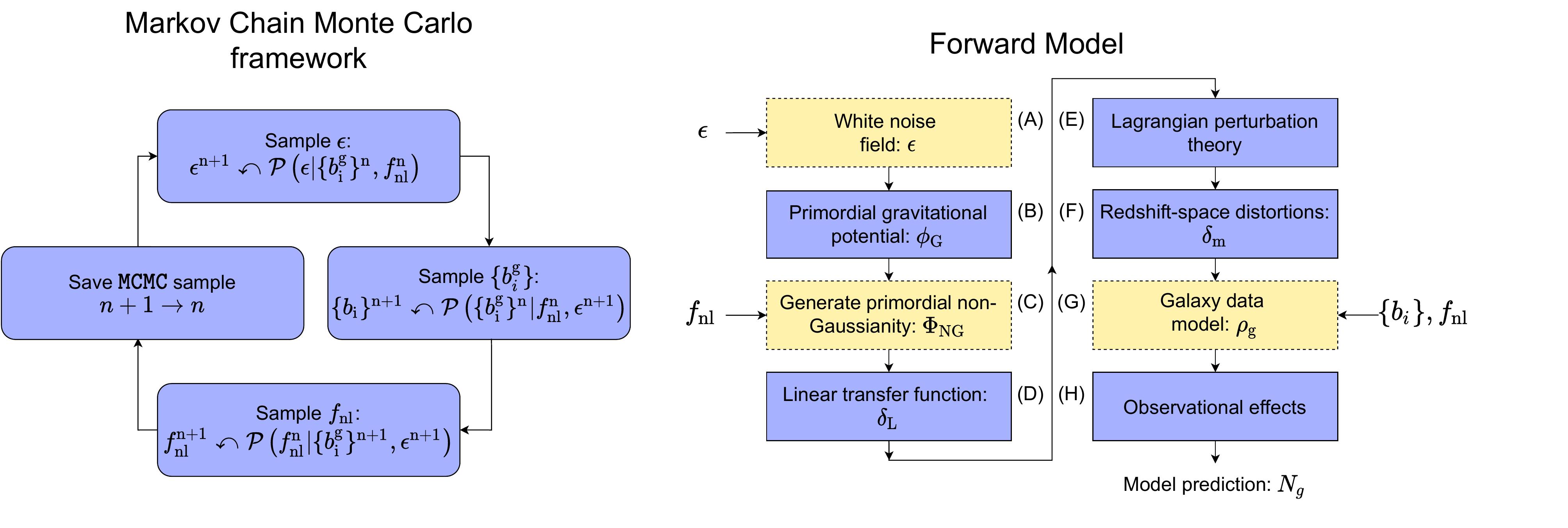}
	\caption{These two flow charts describe the overall process of the \borg{} algorithm. On the left side, we have the Bayesian hierarchical framework, which illustrates the parameter sampling steps. For each sample generated, one full cycle is performed. For more information on the individual sampling schemes, see Section \ref{sampling_steps}. For each parameter evaluation, the algorithm runs one full cycle of the forward model, which is depicted on the right side of the plot. This flow chart illustrates the process of bringing a set of initial conditions (the white noise field $\epsilon$) to a prediction for the expected galaxy field. The likelihood is then computed (as in equation \ref{eq:gaussian_LH}). In this way, the maximum amount of information is conserved in the data, since the algorithm can iteratively fit the field to the data. The yellow, dashed regions are the stochastic phases of the algorithm, which are probabilistically explored in the sampling steps, while the blue, lined regions are the deterministic regions; here, the algorithm transforms the input field accordingly. The letter (A)->(H) are references to more detailed descriptions in the text.\protect\footnotemark}
	\label{fig:flowchart}
\end{figure*}

\section{Physical model and statistical method}
\label{Method}

In the present work, we rely on the \borg algorithm to run a Markov chain which produces Monte Carlo samples of the joint posterior distribution of the 3D initial conditions, the galaxy bias parameters, and constraints on primordial non-Gaussianity (as parameterised by the local form). This section describes the implementation of the Bayesian forward modelling approach to infer the PNG signal from galaxy data, starting with a condensed overview of the \borg{} algorithm. Thereafter, we describe the necessary modifications of its data model to account for the phenomenology of PNG. Specifically, we will include the generation of primordial matter fluctuation fields (equation \ref{eq:fnl_pert}) and a scale-dependent bias (equation \ref{eq:galbias_formula}) to relate observed galaxies to the underlying matter distribution. 

\subsection{The \borg{} algorithm}

The \borg algorithm is a Bayesian hierarchical inference framework aiming to analyse the
cosmic structure by physical forward modelling three-dimensional galaxy fields in cosmological surveys \citep[]{jasche_bayesian_2013,2015JCAP...01..036J,2016MNRAS.455.3169L,ramanah_cosmological_2019,2019A&A...625A..64J}. It uses models of gravitational structure formation
to link the observed distribution of galaxies to the initial three-dimensional density field. In this way, the algorithm re-formulates the inverse problem of inferring the large-scale structures into an initial conditions problem. The naive procedure would be the following: start by proposing a set of initial conditions which is physically forward modelled to an observable galaxy field. First, the gravitational evolution of the linear matter field is simulated by a structure formation process. This yields physical realisations of the underlying dark matter density field in the present Universe. Secondly, the evolved density field is populated through a galaxy data model. Through this process, one would generate the model predictions which are compared to observed galaxy redshift data. A simplified overview of the algorithm for this paper, is given in Fig. \ref{fig:flowchart}. \borg's ability to infer the primordial matter fluctuation field from galaxy surveys is key to the present work.

\footnotetext{For a description on (A) to (D), see section \ref{A_D}; for a description on (E) to (F), see section \ref{E_F}; for a description on (G), see section \ref{G}; for a description on (H), see section \ref{H}.}

The \borg algorithm uses a combination of Hamiltonian Markov chain Monte Carlo and slice sampling techniques to explore the joint posterior of the data \citep[]{2012MNRAS.425.1042J,2019A&A...625A..64J}. This includes the full three-dimensional initial density fields, cosmological parameters and nuisance parameters of the data model, such as galaxy bias parameters. The algorithm performs a statistically rigorous analysis by executing an iterative Markov chain. The primordial initial conditions, including PNG, are combined with the gravitational structure formation model to generate model predictions. In this case, model predictions are the galaxy number counts in individual volume elements. These galaxy number counts are then compared to the observed galaxy distribution using a likelihood estimation, which accounts for the observational noise and the galaxy bias. For every model algorithm iteration, the $\fnl$ parameter, the bias parameters, and initial conditions are updated. The iteration of these processes yields a valid Markov chain permitting us to quantify the significance of inferred quantities \citep[]{jasche_bayesian_2013}.

As outlined, the \borg{} algorithm permits us to infer the three-dimensional initial conditions out of which the present structures formed. To achieve optimal extraction, we will use a new forward model to jointly infer the three-dimensional initial white-noise field $\epsilon$ and the non-Gaussianity parameter $\fnl$, conditioned on the galaxy number counts data $N_{\textrm{g}}^\textrm{o}$,with $\textrm{g}$ indicating the galaxy catalogue index, and $\textrm{o}$ that it is an observed quantity. The vector $\epsilon$ describes the phases of the cosmic structure, while $\fnl$ is introduced in equation \eqref{eq:fnl_pert}. Formally, the posterior $\pi(\epsilon, \fnl, \{b_i \}|N_{\textrm{g}}^\textrm{o})$ can be written as:
\begin{eqnarray}
\pi(\epsilon, \fnl, \{b_i \}|N_{\textrm{g}}^\textrm{o}) = \pi(\fnl)\, \pi(\epsilon)\, \pi(\{b_i^\textrm{g} \}) \frac{\pi(N_{\textrm{g}}^\textrm{o}|\epsilon, \fnl, \{b_i \})}{\pi(N_{\textrm{g}}^\textrm{o})} \, , 
\label{eq:full_posterior}
\end{eqnarray}
where $\pi(\epsilon)$ is the Gaussian white-noise prior with zero mean and unit standard variance, $\pi(N_{\textrm{g}}^\textrm{o}|\epsilon, \fnl)$ is a likelihood distribution, and $\{b_i \}$ are the bias parameters. It is important to remark that in this formulation, the entire data modelling is expressed within the likelihood distribution, which includes the generation of initial conditions and the subsequent forward modelling steps.

\begin{table}[t]
\begin{tabular}{|ccccc|}
\hline
 \# & \begin{tabular}[c]{@{}c@{}}Survey \\ mask\end{tabular} & \begin{tabular}[c]{@{}c@{}}Box size \\ \Mpch \end{tabular}&\begin{tabular}[c]{@{}c@{}}Resolution \\ \Mpch\end{tabular}&$f_{\mathrm{nl}}^\mathrm{true}$  \\ \hline
1      & \footnotesize{\sdssiii{}}       & 4000                                                                          & 31.25                                                                                            & 5                             \\
2      & \footnotesize{\sdssiii{}}       & 4000                                                                          & 15.625                                                                                           & 5                              \\
3      & \footnotesize{\textit{Stage IV}}      & 8000                                                                          & 62.5                                                                                             & 5                              \\
4      &  \footnotesize{\textit{Stage IV}} & 8000                                                                          & 31.25                                                                                            & 5                              \\
5      & \footnotesize{\textit{Stage IV} (Fixed bias)}    & 8000                                                                          & 31.25                                                                                           & 5                              \\ \hline
\end{tabular} \\
{\footnotesize{ The main run of this paper is run \#4. The other runs serve as complementary runs, which have the purpose of investigating the impact of resolution, performance of the method on \sdssiii{}-like data, to demonstrate the reliability of our method, and the effect of fixing the bias values.}}
\caption{An overview of the runs performed in this work.}
\label{tab:runs}
\end{table}

\subsection{Forward modelling large-scale structures with primordial gravitational potential} 
\label{fnl_perturbation}

\begin{figure*}
	\center
    \includegraphics[width=2.0\columnwidth]{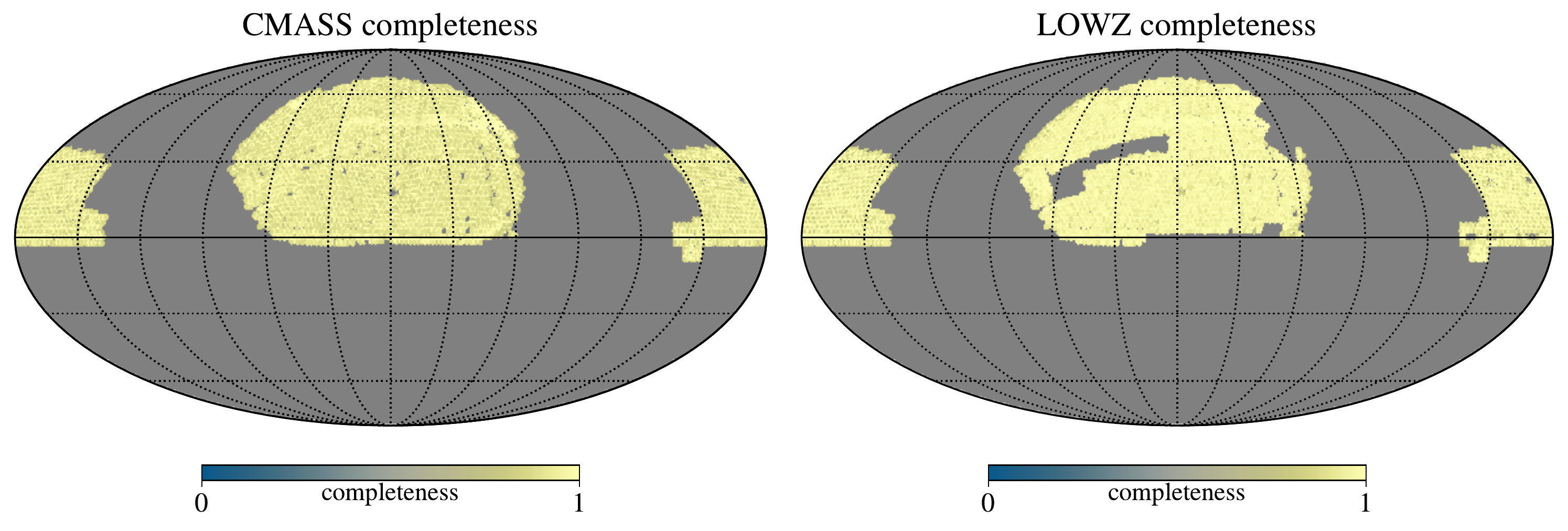}
	\caption{The sky maps displaying the observed and masked regions for the \sdssiii-like runs of this study. In the left, we have illustrated the CMASS component, while in the right we have illustrated the LOWZ component. \citep[][]{ross_clustering_2017}}
	\label{fig:galaxy_masks}
\end{figure*}

In this section, we will discuss the required modifications to the physical model used in \borg to account for primordial non-Gaussianity.

\subsubsection{Generating non-Gaussian initial conditions}
\label{A_D}

In this study, we are focused on inferring the local primordial non-Gaussianity at the leading order, expressed through the $\fnl$ parameter, using a forward modelling approach.

To describe the fluctuations of $\phi_{\textrm{G}}$, we convolve the Gaussian white noise field $\mvec{\epsilon}$ with a first transfer function $T_G(k)$. That transfer function is designed such that $\phi_{\mathrm{G}}$ has a correlation structure given by a primordial power spectrum. In this work, we use the classical parameterisation of the primordial power spectrum in terms of $A_s$ and $n_s$. In the discrete case, the covariance of the field $\phi_\mathrm{G}$ becomes as follows:
\begin{eqnarray}
    \langle \hat{\phi}_{G,\mvec{a}} \hat{\phi}^{*}_{G,\mvec{b}} \rangle = V \delta^K_{\mvec{a},-\mvec{b}} A_{\textrm{s}} \frac{2\pi^2}{k_\mvec{a}^3} \left(\frac{k_\mvec{a}}{k_\mathrm{pivot}}\right)^{n_s-1} \equiv \delta^K_{\mvec{a},-\mvec{b}} T_G(k_\mvec{a})^2,
\end{eqnarray}
with $V$ the volume of the box holding the forward modelled universe, $\mvec{a}$ and $\mvec{b}$ mesh indices, and $\delta^K_{\mvec{a},\mvec{b}}$ the Kronecker delta. $A_{\textrm{s}}$ provides the amplitude of adiabatic, scalar, fluctuations of the potential in the limit $\fnl=0$. Thereafter, we generate the real space primordial gravitational potential as:
\begin{eqnarray}
\centering
    \phi_G(\mvec{x}) = \frac{1}{V} \sum_{\mvec{a}} \exp(i \mvec{k}_{\mvec{a}} \cdot \mvec{x}) \hat{\phi}_{G,\mvec{a}},
\label{eq:wn_to_phi}
\end{eqnarray}
with $\mvec{a}$ a mesh index running over the 3D modes of the box, $\hat{\phi}_{G,\mvec{a}}$ the Fourier amplitude of the modes of the Gaussian random field $\phi_G(x)$. 

In this model, a non-Gaussian primordial Bardeen potential $\Phi_{\textrm{NG}}(\mvec{x})$ is generated from a given Gaussian field $\phi_{\textrm{G}}(\mvec{x})$ through the following Taylor expansion, as highlighted by the box (C) in Figure \ref{fig:flowchart}:
\begin{eqnarray}
\label{eq:fnl_pert}
\Phi_{\textrm{NG}}(\mvec{x}) = \phi_{\textrm{G}}(\mvec{x}) +  \fnl \left (\phi_{\textrm{G}}(\mvec{x})^2 - \langle \phi_{\textrm{G}}(\mvec{x})^2 \rangle  \right) \,.
\end{eqnarray}
The parameters $A_s, n_s$, and $\fnl$ can be related to the inflaton potential \citep[][]{1990PhRvD..42.3936S,2000MNRAS.313..141V,2001PhRvD..63f3002K}.

Once the Bardeen potential is calculated, one can follow the usual modelling techniques to set up the initial conditions to start evolving large-scale structures \citep{peebles_large-scale_1980,1984ApJ...285L..45B,Crocce_2006}. As the evolution at high redshift is linear, the relation between matter density $\hat{\delta}_{\mathrm{L}}(\mvec{k})$ and the primordial potential is, as highlighted by the box (D) in Figure \ref{fig:flowchart}:
\begin{eqnarray}
    \hat{\delta}_{\mathrm{L}}(\mvec{k},z) = T(k,z) \Phi_{\textrm{NG}}(\mvec{k})\,,
\label{eq:lin_dens_from_phi}
\end{eqnarray}
where $z$ is taken at an arbitrary reference epoch. The transfer function $T(k,z)$, which contains the linear growth factor, is provided either by CLASS \citep{Lesgourgues_2014} or through analytic computation \citep{EH98,EH99,takada_2006}.

\subsubsection{Simulating structure growth}
\label{E_F}
For this proof-of-concept, we construct the matter density field at present time by  relying on the first-order Lagrangian Perturbation Theory (LPT), commonly known as Zel'dovich approximation. To compute the present-day density field, we populate the Lagrangian coordinates $\mvec{q}$ with a set of particles that are displaced according to:
\begin{eqnarray}
    \mvec{\Psi}(\mvec{q},z) = - \mvec{\nabla}\Delta^{-1} \delta_{\mathrm{L}}(\mvec{q},z)\,.
\end{eqnarray}
The particle at the position $\mvec{x}=\mvec{q}+\mvec{\Psi}(\mvec{q},z)$ is then assigned, with the Cloud-In-Cell kernel, to a regular grid, as highlighted by the box (E) in Figure \ref{fig:flowchart}. In this work, we evaluate at $z=0$.

As a final step, we apply redshift-space distortion effects to the particles, by transforming them from their rest frame to the comoving frame as highlighted by the box (F) in Figure \ref{fig:flowchart}. This incorporates both the Kaiser and \textit{Fingers of God} effects, to the extent that they are captured by first-order LPT, which may impact the inference of $\fnl$ \citep[][]{di_dio_non-gaussianities_2017,Tellarini_2016,bharadwaj_quantifying_2020,Tellarini_2016,Karagiannis_2018}. Handling RSDs allows us to include the peculiar velocities when constraining $\fnl$ with our field-based approach.

Although LPT is sufficient to describe large-scale features, small-scale features suffer from approximation issues. In this work, we focus on scales of $\geq 16~\Mpch$, which is where LPT is a good approximation. It should be noted that the implementation described here allows us to further improve and modify the forward model. As such, a more accurate structure formation model would allow the method to go beyond the current resolution limit. In particular, other structure growth models exist which can provide us with results including more physical effects, e.g., the particle mesh model \citep[][]{2019A&A...625A..64J} or second- \citep[2LPT;][]{jasche_matrix-free_2015} and higher-order LPT (\cite{paper_nLPT}). Thus, while in this proof-of-concept we are relying on LPT to describe structure formation, we can improve the structure formation model to provide more realistic, nonlinear density field realisations. Moreover, this allows us to combine the $\fnl$ perturbation model with other cosmological models, e.g., a dark energy model \citep[as done in][]{ramanah_cosmological_2019}. More information about this modular approach, of which the $\fnl$ perturbation model is a part of, can be found in the literature \citep[see][]{ramanah_cosmological_2019}.

\subsection{Scale-dependent galaxy biasing}
\label{G}

The next step of the forward modelling is to model the relationship between the predicted matter field and the observation of galaxy number counts. In the field of large-scale structure cosmology, galaxies are considered tracers of the underlying gravitational potential. Specifically, galaxies are \textit{biased} tracers, displaying clustering properties approximately align with the underlying dark matter \cite[][]{kaiser_spatial_1984}. The uncertainty of this model systematic effect is currently one of the most important unresolved issues in LSS cosmology, hindering the nonlinear analysis of galaxy surveys. For a review of the galaxy bias problem, the interested reader is referred to the literature \citep{desjacques_large-scale_2018,schmidt_rigorous_2019}.

 At linear order, the relationship is the linear bias function: $\rho_{\textrm{g}} = \langle N_{\textrm{g}}^{\mathrm{o}} \rangle(1+b_1^{\mathrm{g}}\delta_{\mathrm{m}})$, where $\rho_{\textrm{g}}$ is the resulting galaxy field, $\langle N_{\textrm{g}}^{\mathrm{o}} \rangle$ is mean number of galaxies, $b_1^{\mathrm{g}}$ is the linear bias and $\delta_{\mathrm{m}}$ is the final density field, in the present-time universe. $\mathrm{g}$ is the index of the galaxy catalogue. In addition to a linear bias, non-Gaussian perturbations imprint a scale-dependent bias effect in dark matter halo formation. This effective parameter describes the effect of the PNG-induced coupling between the long-wavelength potential perturbations and the short-wavelength modes of the density field on galaxy formation \citep[]{dalal_imprints_2008,Slosar_2008,Desjacques_2009}. While the scale-dependence of this bias is known, its amplitude $b_\phi$ depends on the details of the given galaxy sample and cannot be predicted in general; see \cite{Pillepich_2009,Biagetti_2017} for measurements of $b_\phi$ on simulated dark matter halos, and \cite{Barreira_2020} for a measurement on simulated galaxies.

Here, as in all published constraints on $\fnl$ so far, we will assume the universality relation for $b_\phi$ \citep[]{dalal_imprints_2008,Slosar_2008}, which relates the amplitude of the scale-dependent bias to the linear bias $b_1^{\mathrm{g}}$.

In terms of the bias contribution: $$\rho_{\textrm{g}} = \langle N_{\textrm{g}}^{\mathrm{o}} \rangle \left \{ 1+ \left [b_1^{\mathrm{g}} + \Delta b(k, \fnl) \right ] \delta_{\mathrm{m}} \right \} \, ,$$ the scale-dependent bias $\Delta b(k, \fnl)$ then takes the following shape:
\begin{eqnarray}
    \Delta b(k, \fnl) = b_{\phi} \alpha(k,z) = b_{\phi} \fnl\frac{3\Omega_{\mathrm{m}} H^2_0}{2 c^2 k^2 T(k,z)}\,.
    \label{eq:b_ng}
\end{eqnarray}
Under the universal mass approximation \citep[]{1974ApJ...187..425P,Sheth_1999,Pillepich_2009}, $b_{\phi}$ takes the form of \citep[]{matarrese_effect_2008,Slosar_2008,Schmidt_2013}: 
\begin{eqnarray}
    b_{\phi} = 2 \delta_{\mathrm{c}}(b_1^{\mathrm{g}}-1) \, ,
    \label{eq:b_phi}
\end{eqnarray}
where $\delta_{\mathrm{c}}$ is the spherical collapse threshold, $\delta_{\mathrm{c}} \approx 1.686$ \citep{1972ApJ...176....1G}. 

\label{linear_sdb}
\begin{table*}
\centering
\begin{tabular}{|lllll||lllll|}
\hline
\sdssiii{}     & $b_1^{\mathrm{g}}$ &  $z_{\mathrm{min}}$/$z_{\mathrm{max}}$ &   $\bar{n}_{\mathrm{gal}}$ ($10^{-4}h^{3}\text{Mpc}^{-3}$)  & \textit{Stage IV}     & $b_1^{\mathrm{g}}$ & $z_{\mathrm{min}}$/$z_{\mathrm{max}}$ & $\bar{n}_{\mathrm{gal}}$ ($10^{-4}h^{3}\text{Mpc}^{-3}$) \\ \hline
Cat. 1 &  3.22       &       0.4/0.75    &  $3.01 $      & Cat. 1 &  1.46     &   0.90/1.10          &   6.86       \\
Cat. 2 &   2.48         &   0.4/0.75    &  $2.99 $    & Cat. 2 &  1.61     &      1.10/1.30       &  5.58       \\
Cat. 3 &  1.78        &   0.2/0.5   & $6.31 $     & Cat. 3 &   1.75    &       1.30/1.50      &    4.21      \\
Cat. 4 & 0.87          &   0.2/0.5    & $10.4 $      & Cat. 4 &  1.90     &      1.50/1.80       &   2.61        \\ \hline
\end{tabular}
\caption{Configuration of tracer catalogues and bias parameters. $b_1$ is the linear bias parameter, $\bar{n}_{\mathrm{gal}}$ is the number density, and $z$ is the redshift distance of the tracer bins. The noise $\sigma_{\textrm{g}}$, introduced in equation \eqref{eq:gaussian_LH}, is fixed to $\sqrt{\langle N_{\textrm{g}}^{\mathrm{o}} \rangle}$, the mean number of galaxies in catalogue $g$.}
\label{tab:configs}
\end{table*}

Thus, the galaxy bias model is formulated with the linear bias $b_1^{\mathrm{g}}$  and with the scale-dependent bias effect $b_{\phi}(k,\fnl)$, which translates the dark matter density contrast field $\delta_{\mathrm{m}}$ to the galaxy field $\rho_{\textrm{g}}$ as highlighted by the box (G) in Figure \ref{fig:flowchart}:

\begin{eqnarray}
    \rho_{\textrm{g}}\left(\delta_{\mathrm{m}},\langle N_{\textrm{g}} \rangle,b_1^{\mathrm{g}},\fnl\right) = \langle N_{\textrm{g}} \rangle \left\{1 + \left[b_1^{\mathrm{g}} + \Delta b\left(k, \fnl\right)\right] \delta_{\mathrm{m}}\right\} .
\label{eq:galbias_formula}
\end{eqnarray}
This is the expression for the galaxy field in a universe with local PNG, to the leading order. We leave the incorporation of higher-order bias terms of local PNG to future work \citep[see e.g.][ for a description of the said higher-order terms]{Assassi_2015,Moradinezhad_Dizgah_2021,Barreira_2020,Barreira_2020b, barreira_local_2021,barreira_predictions_2021}. Notice also that the scale-dependent bias should strictly involve the primordial potential at the \textit{Lagrangian} (initial) position, rather than the late-time Eulerian position \citep{Assassi_2015}. The difference is, however, only a second-order effect.

Finally, when constraining $\fnl$ from galaxy clustering, it is in general
important also to include the so-called relativistic effects \citep[see][for a review]{jeongLargescaleStructureObservables2015}. We leave the incorporation of these effects into the forward model to future work.

\subsubsection{A likelihood model of galaxy data}
\label{H}

\begin{figure}
	\center
    \includegraphics[width=1.0\columnwidth]{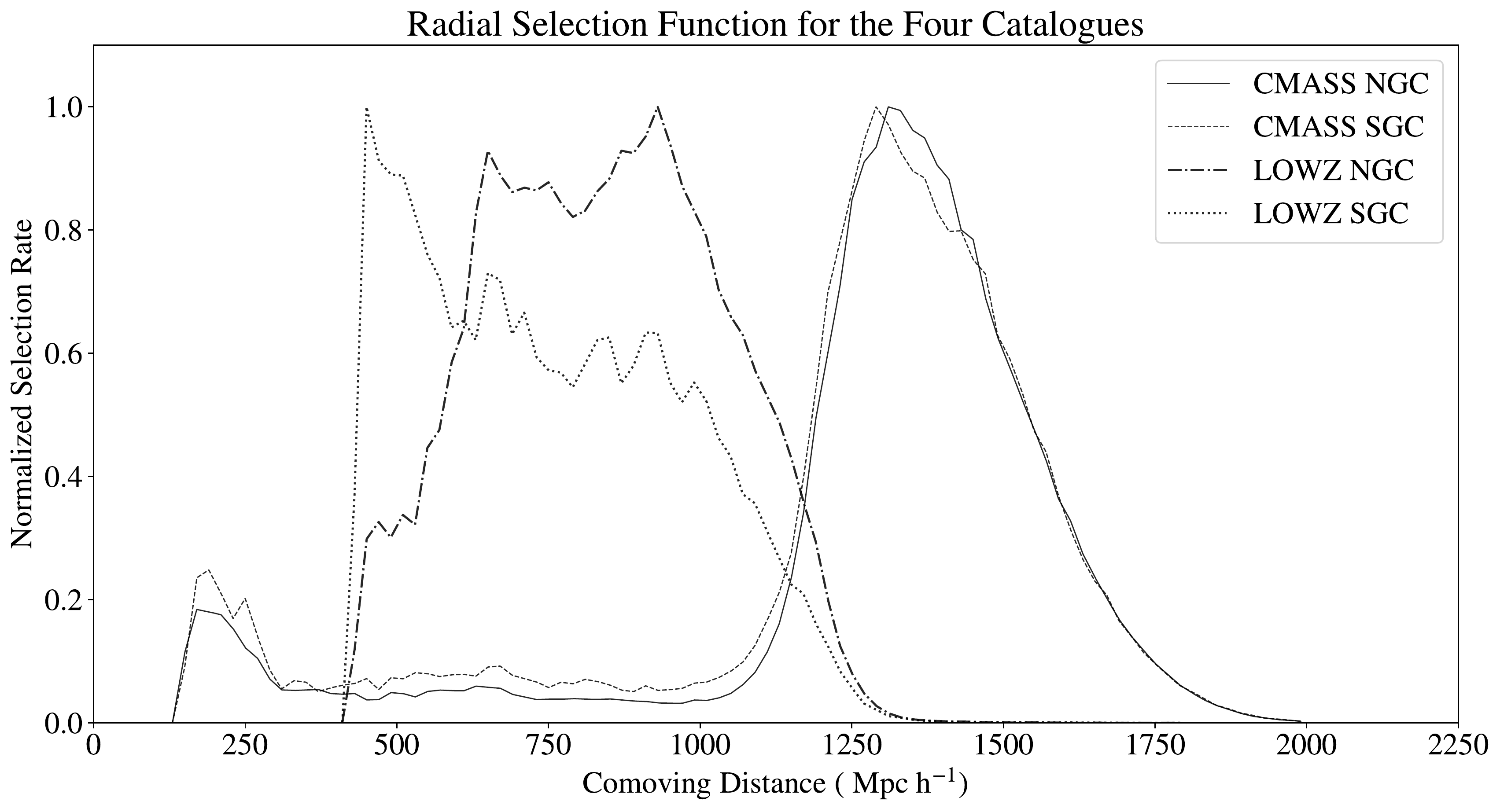}
	\caption{The radial selection functions for the \sdssiii-like runs of this study. The solid and dashed lines depict the function for the two galaxy catalogues included in the CMASS sample, the North and the South Galactic Cap (NGC, SGC), while the dash-dotted and dotted lines describe the function for the two galaxy catalogues in the LOWZ component, NGC and SGC. \citep[][]{ross_clustering_2017}}
	\label{fig:radial_sel}
\end{figure}

With a prediction of $\rho_{\textrm{g}}$, the next step is to relate this field to $N_{\textrm{g,p}}$, the expected number of observed galaxies at a given voxel $\mathrm{p}$. Since the galaxies are split into several catalogues, each with different survey geometries and nuisance parameters, we denote the catalogue for each quantity with $\mathrm{g}$. In this way, by applying $R_{\textrm{g,p}}$, the linear response operator of the survey, we obtain as highlighted by the box (H) in Figure \ref{fig:flowchart}:
\begin{eqnarray}
    N_{\textrm{g,p}} = R_{\textrm{g,p}} [\rho_{\textrm{g,p}}(\delta_{\mathrm{m}},\langle N_{\mathrm{g}} \rangle ,b_1^{\mathrm{g}},\fnl)] \, ,
\label{eq:dens_to_lambda}
\end{eqnarray}
where $\langle N_{\mathrm{g}} \rangle$ is the mean number of galaxies in catalogue $\mathrm{g}$. The purpose of $R_{\textrm{g,p}}$ is to account for survey geometry and selection effects. We emphasise that the galaxy data model accounts for the fact that galaxies are biased tracers of the underlying density field, as well as for the discrete nature of galaxy formation \citep[]{layzer_new_1956,peebles_large-scale_1980}.

As for the noise of the galaxy counts, in this work, we are at sufficiently large scales to describe it as a Gaussian likelihood \citep[][]{Kitaura_gaussian,jasche_bayesian_2010,jasche2017}. For a galaxy catalogue $\textrm{g}$, the log-likelihood is given as:
\begin{multline}
    \mathrm{ln} \left [ \pi\left (  N_{\textrm{g,p}}^{\mathrm{o}} \, \big| \, N_{\textrm{g,p}} \, , \sigma_{\textrm{g}} \right ) \right ] = -\textrm{P} \, \mathrm{ln} \left [\sqrt{2\pi} \, \sigma_{\textrm{g}} \right ] - \\ \frac{1}{2}\sum_{\textrm{p}=0}^{\textrm{P-1}} \left ( \frac{N_{\textrm{g,p}}^{\mathrm{o}}-N_{\textrm{g,p}}}{\sigma_{\textrm{g}}} \right ) ^2 \, ,
\label{eq:gaussian_LH}  
\end{multline}
where $N_{\textrm{g,p}}^{\mathrm{o}}$ is the observed galaxy number count in voxel $\mathrm{p}$, and $N_{\textrm{g,p}}$ is the model prediction of the physical forward model, given $\epsilon$ and $\fnl$. $\textrm{p}$ is the voxel index, and $\textrm{P}$ is the total number of voxels. $\sigma^{g}$ is the noise of the galaxy catalogue; in this work, this is fixed to the mean number of galaxies in catalogue $g$:
\begin{eqnarray}
    \sigma_{\textrm{g}}=\sqrt{\langle N_{\textrm{g}}^{\mathrm{o}} \rangle} = \sqrt{\frac{\Sigma_\textrm{p} N_{\textrm{g,p}}}{\Sigma_\textrm{p} R_\textrm{g,p}}},
\end{eqnarray}
where $R_\textrm{p}$ is the selection function for each voxel; summing over $R_\textrm{p}$ yields the total number of observed voxels.
This is chosen such as to emulate Poisson noise. 
A comment on the likelihood model: while the uncertainty on galaxy counts is usually assumed to be Poisson distributed \citep{layzer_new_1956,peebles_large-scale_1980,ramanah_cosmological_2019}, in this study, we assume that it is Gaussian distributed. The justification for this is that we have a low resolution in the voxels, resulting in a relatively high number of galaxies per voxel. As the Poisson distribution approaches a Gaussian distribution in the limit of large values, we argue that this choice has little effect on the outcome of the analysis \citep{2021JCAP...03..058N}.

In this way, we obtain a likelihood distribution that contains the statistical process of generating galaxy observations given a sampled set of initial conditions. Hence, due to the deterministic nature of structure formation, the expected number of galaxies per voxel $N_{\textrm{g,p}}$ is related to the white-noise field $\epsilon$ via the physical forward model (described in Section \ref{fnl_perturbation}) and the galaxy bias model (discussed in Section \ref{linear_sdb}). 

\subsection{Sampling of density field, bias parameters, and $\fnl$}
\label{sampling_steps}

Within the framework of \borg{}, the main objective is to explore the full, joint posterior distribution of the cosmic density field, bias parameters, and $\fnl$. However, directly sampling from the joint posterior is difficult, possibly numerically infeasible. Instead, \borg{} evaluates the conditional posterior distributions separately, to sample new proposals for the white-noise field $\epsilon$, galaxy bias parameter $\{b_{\mathrm{i}}^{\textrm{g}} \}$ (where $g$ is the galaxy catalogue number, and $i$ is the index for the galaxy bias parameter), and $\fnl$. In this way, the conditional posteriors can be sampled from in a sequential manner. This block sampling strategy is outlined in the following way:

\begin{eqnarray}
	(1) &\: \epsilon^{\mathrm{n}+1} \curvearrowleft \pi \left(\epsilon |   \{b_{\mathrm{i}}^{\textrm{g}}\}^\mathrm{n}, f_{\mathrm{nl}}^{\mathrm{n}} \right) \, , \nonumber \\
	(2) &\: \{b_{\mathrm{i}}^{\textrm{g}}\}^{\mathrm{n}+1} \curvearrowleft \pi \left(\{b_{\mathrm{i}}^{\textrm{g}}\} |   f_{\mathrm{nl}}^{\mathrm{n}}, \epsilon^{\mathrm{n}+1}  \right) \, , \\
	(3) &\: f_{\mathrm{nl}}^{\mathrm{n}+1} \curvearrowleft \pi \left(f_{\mathrm{nl}}^{\mathrm{n}} | \{b_{\mathrm{i}}^{\textrm{g}}\}^{\mathrm{n}+1},  \epsilon^{\mathrm{n}+1}\right) \, ,\nonumber 
\label{eq:block_sampling}
\end{eqnarray}
where $\curvearrowleft$ represents the process of drawing a value from the probability distribution, and $n$ is the index of the sample. By consecutive sampling one parameter of the physical model at a time, the conditional distributions form a sample of the full joint posterior distribution \citep[]{hastings_1970}. See the left-hand side of Figure \ref{fig:flowchart} for a graphical representation of the sampling schemes.

For the field $\epsilon$, the main problem consists of sampling a high-dimensional parameter space, with each dimension consisting of the amplitude in each voxel. Utilising standard Markov chain Monte Carlo methods yields insufficient results; mostly due to the numerical intensity of each forward model calculation, coupled with a high rejection rate. Instead, we apply a Hamiltonian Monte Carlo algorithm which guides the proposal procedure of new samples by utilising dynamical physical symmetries. For a more detailed description of the sampling of the white-noise field $\epsilon$, see previous works on the subject \citep{jasche_bayesian_2013,ramanah_cosmological_2019}.

In this study, the posterior distributions of $\fnl$ and galaxy bias parameters are probed using a slice sampling procedure \citep{neal_markov_2000,neal_slice_2003}, which has a unit acceptance rate. For each realisation of the density field, the algorithm samples the posterior distribution of $\fnl$, and accepts a new value for $\fnl$ for the next iteration. We use a prior with standard deviation set to $100$. In the same fashion as for the $\fnl$ parameter, the galaxy bias parameters are sampled and updated by a slice sampling approach.

\begin{figure*}
	\center
	\includegraphics[width=2.0\columnwidth]{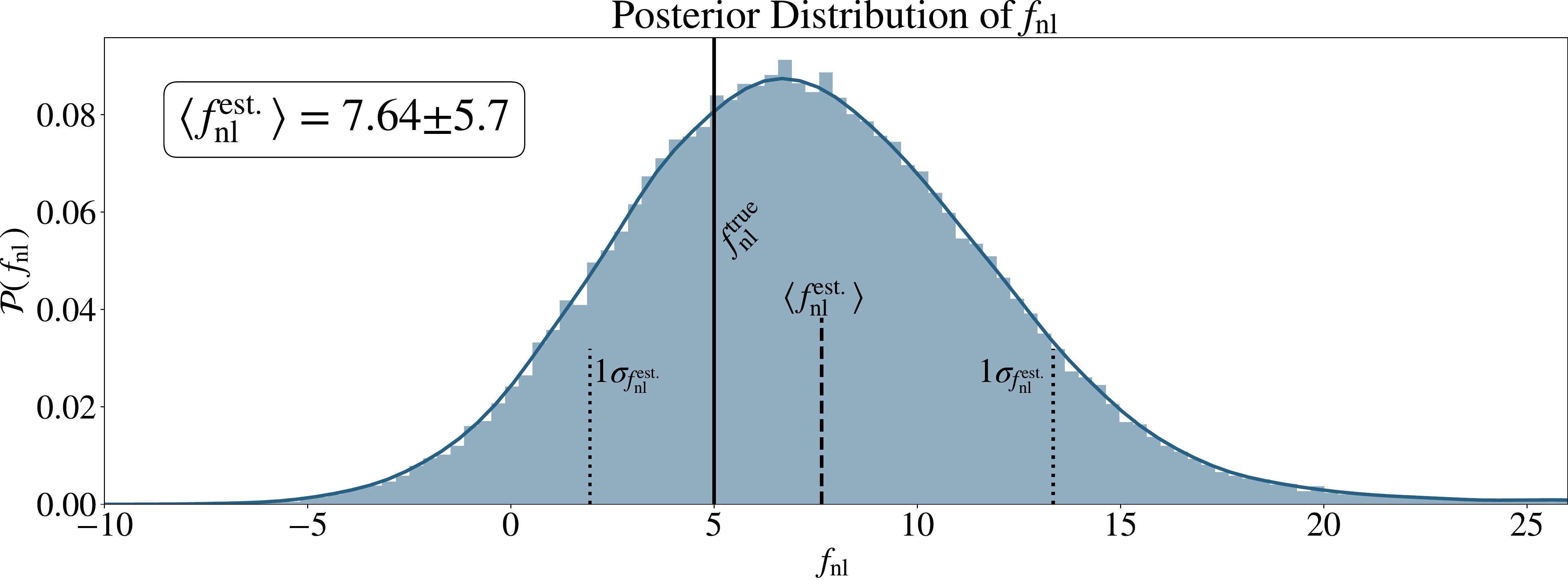}
	\caption{The $\fnl$ values sampled of run \#4: the high resolution \textit{Stage IV} data set. The figure displays the conditional posterior distribution of $\fnl$, for the given mock data. Notice how the fiducial $\fnl$ value is situated within 1$\sigma$ of the sampled mean.}
	\label{fig:trace_stage_iv}
\end{figure*}

\section{Generating artificial test data}
\label{data_section}

In this section, we describe the mock data sets used to test our method. We are particularly interested in testing the method's ability to constrain $\fnl$ when facing survey geometries and selection effects. Generated mock data sets will be designed to emulate the existing \sdssiii{} survey \citep[]{alam_eleventh_2015} and the future \textit{Stage IV} surveys \citep{lsst_science_collaboration_lsst_2009,2018LRR....21....2A,dore_cosmology_2014}.

\subsection{Creating mock galaxy surveys}
\label{mock_observations}

To generate mock data, we will follow similar procedures as described in previous works, using the physical data model as outlined in Section \ref{Method} \citep[]{jasche_fast_2010,jasche_bayesian_2013,ramanah_cosmological_2019}:

\begin{enumerate}
    \item For the sake of demonstration, we aim to evaluate the physics forward model on a cubic Cartesian box of side length $L=4000 \, \Mpch$ (\sdssiii{}-like) or $L=8000 \, \Mpch$  (\textit{Stage IV}) and $N_{\text{grid}}=128$ or $256$. This yields grid resolutions in the range of $\Delta L\simeq 62.5\Mpch$ to $\Delta L\simeq 15.6$.
    \item A random three-dimensional field $\epsilon$, with zero mean and unit standard deviation, is generated. Given this white-noise field, we compute a primordial density field by applying the primordial power spectrum, perturbing it with the $\fnl$ parameter, and by applying the cosmological transfer function. This yields the linear matter field $\delta_L$, which is the starting point for the gravitational structure formation model. This step follows equations \eqref{eq:wn_to_phi} to \eqref{eq:lin_dens_from_phi}.
    \item Although the \borg algorithm permits running full gravitational particle mesh simulations, at this coarse resolution Lagrangian perturbation theory provides a viable approximation to the structure formation problem \citep{moutarde_precollapse_1991,buchert_testing_1993,bouchet_perturbative_1994,scoccimarro_bispectrum_2000,scoccimarro_pthalos_2002}. To reduce the noise of the simulated particle distribution, we oversample the initial density by a factor $2$ yielding a total number of $(2N)^3$ simulation particles. Particles are then evolved to the present epoch, using LPT, and are assigned to a three-dimensional Cartesian grid via the Cloud-In-Cell (CIC) kernel to yield the evolved three-dimensional density field. In addition, we add the redshift-space distortions to transform the particles from rest frame to redshift frame.
    \item To emulate a biased galaxy distribution, we apply the scale-dependent galaxy bias (described in Section \ref{linear_sdb}) to the forward modelled density field, to obtain the galaxy field.
    \item Finally, we apply the radial selection functions and survey geometry to the simulated galaxy field to emulate the observational effects of the respective surveys detailed below.
\end{enumerate}

An overview of the runs and the data sets are organised in Table \ref{tab:runs} and Table \ref{tab:configs}, together with the detailed parameter choices for the galaxy bias model. For a more detailed description of the \sdssiii{} mock data, see Section \ref{sdss3}. For a more detailed description of the \textit{Stage IV} mock data, see Section \ref{stage_iv}. 

To calculate the cosmological power spectrum and transfer functions, we assume the following set of cosmological parameters ($\Omega_{\mathrm{m}}=0.3111$, $\Omega_{\Lambda}=0.6889 $, $\Omega_{b}=0.0490$, $h=0.6766$, $\sigma_8=0.8102$, $n_s=0.965$) which are taken to be the recent values of \textit{Planck} + \textit{BAO} \citep[][]{planck_2018}. To calculate the required cosmological transfer function we use the prescription provided by \cite{EH98} and \cite{EH99}. 

The mock data sets are generated with $f_{\mathrm{nl}}^{\mathrm{fiducial}} = f_{\mathrm{nl}}^{\mathrm{true}} = 5$. To reiterate, the main aim of this study is to recover this $\fnl$ value and to estimate the accuracy of our method.

\subsection{Emulating \sdssiii{} data}
\label{sdss3}

\begin{figure}
	\center
	\includegraphics[width=1.0\columnwidth]{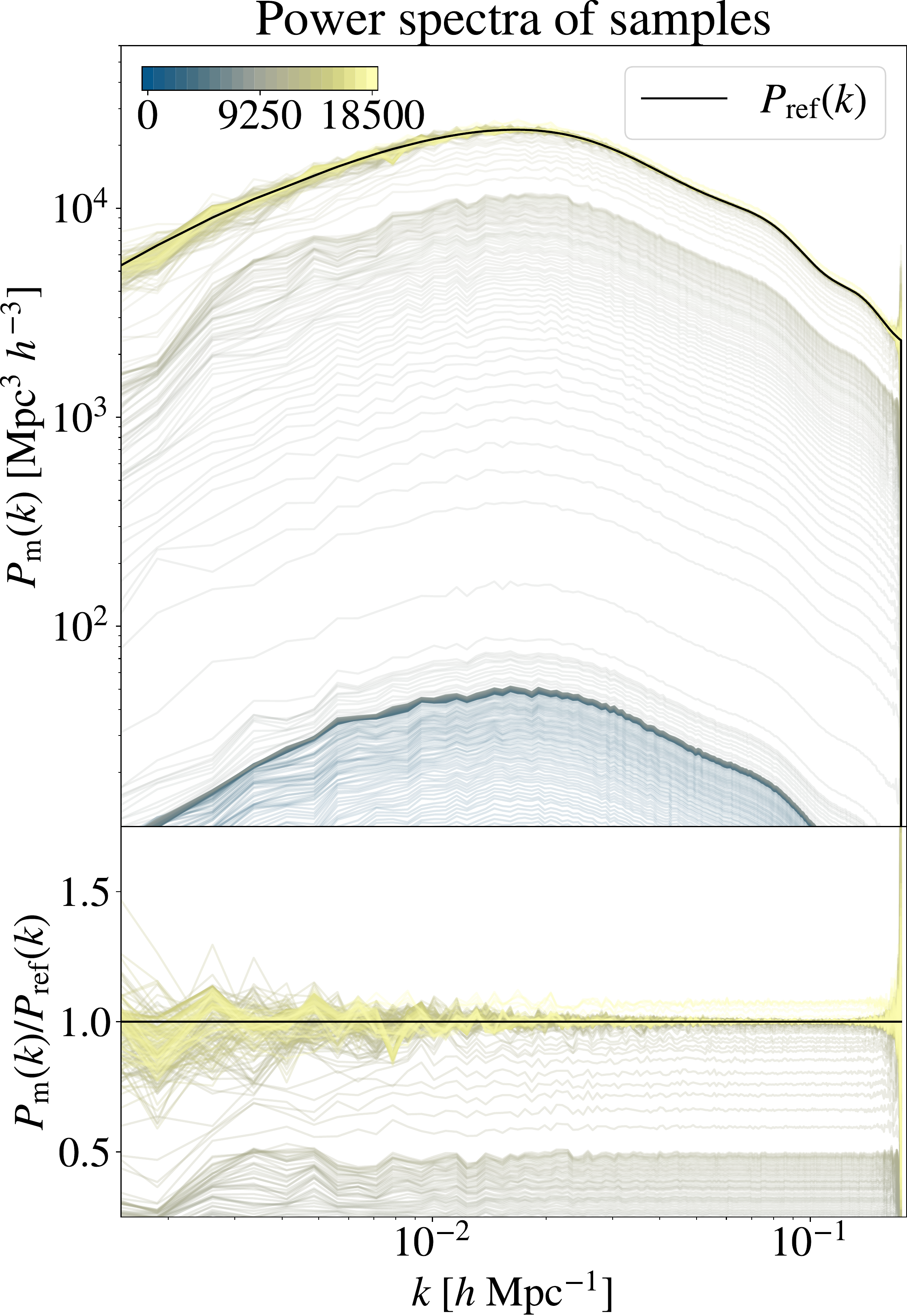}
	\caption{The plot shows the systematic drift of present-time power spectrum amplitudes during burn-in phase of run \#4. Different colours of lines denote measurements at different sequential sampling steps in the Markov chain as indicated by the colour bar. It can be seen that the Markov chain initially starts at a remote place in parameter space but approaches the target region in the parameter space after about $\sim 7000$ Markov chain transitions. Note that these power spectra have been computed during the post-production process; the forward model is independent of the matter power spectra. Thus, the retrieval of a correct power spectra implies that our method is able to retrieve the fiducial cosmology in the input data.}
	\label{fig:burnins}
\end{figure}

We seek to generate mock galaxy surveys emulating the major features of the combined CMASS and LOWZ galaxy samples of the \sdssiii{} survey. We will account for the difference in sky completeness and selection effects by dividing the survey into four sub-catalogues: CMASS Northern Galactic Cap (NGC), CMASS Southern Galactic Cap (SGC), LOWZ NGC, and LOWZ SGC. These sky maps are illustrated in Fig. \ref{fig:galaxy_masks}. Corresponding radial selection functions have been estimated numerically by binning the equivalent distribution of galaxies $N(d)$ in the actual CMASS and LOWZ survey \citep[][]{ross_clustering_2017}, where $d$ is the comoving distance from the observer. The results are plotted in Fig. \ref{fig:radial_sel}. We note that, by jointly analysing four galaxy sub-catalogues, we will perform a multi-tracer analysis. As has been previously demonstrated, using multi-tracer populations reduces the impact of cosmic variance when inferring $\fnl$ \citep{seljak_measuring_2009}. However, since the galaxies are placed in separated sub-catalogues, we neglect any correlated noise between the galaxy sub-catalogues. The specific parameter settings for galaxy biases and the total number of galaxies for the four sub-catalogues are given in Table~\ref{tab:configs}.

The value of expected galaxy number densities $\bar{n}_{\mathrm{gal}}$ for each catalogue have been selected as table~1 in \cite{Alam_2017}. The expected galaxy number density is related to the observed number count as: $\bar{n}_{\mathrm{gal}} = \langle N^{\mathrm{o}}_{\mathrm{g}}\rangle/V_{\mathrm{o}} $, where $V_{\mathrm{o}}$ is the volume containing observed galaxies. The linear biases $b_1^{\mathrm{g}}$ for each tracer catalogue has been selected as per figure~8 in \cite{Lavaux2019}, with light-cone effects adjusted for. 

As a final note, for the tracer populations, we are only including the galaxies of the \sdssiii{} survey, and the observed quasar populations remain outside of the scope of our data cube. Hence, we are probing smaller volumes compared to other works \citep{leistedt_constraints_2014,castorina_redshift-weighted_2019,mueller2021clustering}.

\subsection{Emulating \textit{Stage IV} data}
\label{stage_iv}

In this study, we apply our method to mock data, which are based on forecasts of next-generation galaxy redshift surveys. In the context of this paper, we dub these mock data sets as \textit{Stage IV}. The purpose of these sets is to emulate typical features of next-generation of surveys. In this way, by inferring $\fnl$ in the \textit{Stage IV} mock data sets, we provide an estimate of how well our method could perform on data from the next generation of galaxy redshift surveys.

For the sky area of the \textit{Stage IV} mock data, we design the mock galaxy survey such as to cover a total area of $\sim$15~000 square degrees. This sky area is spread evenly over the north and south poles of the full sky, resulting in a uniform completeness for $|b| > 39^{\circ}$. In other words, we exclude observations in regions $\pm 39^{\circ}$ from the galactic plane. The resulting survey geometry is then represented as a full sky HEALPix map with $n_{\text{side}}=512$ to be used for mock data generation \citep{gorski_healpix_2005}.

In the case of the radial selection function, we split the mock data into 4 galaxy catalogues, each fully covering a redshift bin as in Table \ref{tab:configs}, with a total redshift range of $0.9< z < 1.8$. In other words, the radial selection function $S_{\textrm{g}}(z)$ for the $g$th galaxy catalogue covering the redshift range $z_{\text{min}}\leq z \leq z_{\text{max}}$ can be written as:
\begin{eqnarray}
S_{\textrm{g}}(z)=
    \left\{
        \begin{array}{rl} 
            1, & \text{if } z \in  [z_{\text{min}}, z_{\text{max}}] \\ 
            0, & \text{otherwise}.
        \end{array}\right.
\label{eq:sel_fcts}
\end{eqnarray}
For the choices of linear bias values and number densities, see Table \ref{tab:configs}.The purpose of these sets is to emulate typical features of next-generation surveys and we take  \textit{Euclid}-like specifications as a representative case from \cite{euclid_collaboration_euclid_2020}. For the configuration of resolution and box size, see Table \ref{tab:runs}. We use these specifications to generate a three-dimensional galaxy distribution that emulates relevant features and systematic effects of the \textit{Stage IV} surveys. We illustrate the radial selection function in Figure \ref{fig:stage_iv_rs}, and the sky maps in Figure \ref{fig:stage_iv_galaxy_masks}; these can be found in the appendix (Section \ref{stage_iv_mask}).

In addition, we also perform a fixed bias test run, using the configurations of the high-resolution \textit{Stage IV} mock data. For this run, we keep the bias parameters of the MCMC-chain fixed to their fiducial values. This design choice is to test the scenario when we have a complete understanding of the galaxy biasing. In other words, with this simplified test case, we investigate how marginalising out bias parameters affect the constraining power of $\fnl$.

\section{Results}
\label{results}

The previous section outlined the generation of artificial galaxy data used to test our method. This section provides an analysis of the results of our method when applied to the described mock data. Specifically, we are interested in estimating the ability of the method to infer the $\fnl$ parameter from the data, which is subject to systematic effects such as survey geometries, selection effects, and noise. A summary of the constrained $\fnl$ values can be found in Table \ref{tab:results}.

\subsection{Testing the MCMC warm-up phase}
\label{tests}

The \borg{} algorithm executes a large-scale Markov chain Monte Carlo by exploring the joint posterior of initial density fields, the $\fnl$, and various nuisance parameters, such as bias parameters or unknown noise levels. To test the validity of the MCMC sampler, we initialise the Markov chain from an overdispersed state; \borg{} will start in a remote region of the probability space, and then coherently drift towards the region of highest probability and explore these probabilistically correctly. More specifically, we initialise our Markov chain with a randomly overdispersed Gaussian initial density field that is scaled down to have an amplitude the tenth of a realistic cosmological density field. The warm-up phase of the sampler can then be monitored by following the coherent drift of the power spectra estimated from sequential Markov samples. This coherent drift of the power at all Fourier modes is demonstrated in Fig. \ref{fig:burnins}. It can be seen that the sequentially estimated power spectra approach the correct fiducial cosmological power spectrum after about $7000$ Markov transitions. 

After passing the warm-up phase, \borg{} correctly explores the parameter space of plausible large-scale structure realisations. This is reflected by the fact that power spectra correctly fluctuate around the fiducial power spectrum. The corresponding scatter reflects both cosmic variance and observational noise. Note that \borg{} recovers unbiased estimates of the primordial power spectrum throughout the entire range of Fourier modes considered in this work. This is because our approach correctly handles the systematic effects associated with survey geometries and selection effects. When unaccounted for, these systematic effects will yield significant erroneous power at large scales, concealing the physical signal of PNG from $\fnl$. This successful test therefore demonstrates that our approach has the potential to study the primordial universe physics on the largest scales in galaxy surveys. We emphasise that the analysis pipeline adopts a pure forward model approach, as illustrated in Figure \ref{fig:flowchart}; thus, by correctly sampling the fiducial power spectra, we demonstrate that our method is able to recover a key observable accurately.

We also perform Gelman-Rubin convergence test of our chains, which they pass. The outline of the test and the corresponding results can be found in Appendix \ref{gr_test}.

\subsection{Inferring the non-Gaussianity parameter $\fnl$}

\begin{table}
\centering
\hspace{20mm}
\begin{tabular}{|llll|}
\hline
Run \# & Description &  Sampled $\fnl$ & $\sigma_{\fnl}$ \\ \hline
1      & \sdssiii{}, low reso. &   10.11              &          11.9       \\        
2      & \sdssiii{}, high reso. &  9.89             &          8.78       \\
3      & \textit{Stage IV} low reso. &  0.59              &       7.09          \\
4      & \textit{Stage IV}, high reso. &   7.64           &      5.70           \\
5      & \textit{Stage IV}, fixed bias, high reso. & 3.17     &       4.43          \\ \hline
\end{tabular}
\caption{Summary of the measured $\fnl$ values for the different configurations. Our method is able to consistently measure the fiducial $\fnl$ value, up to various degrees of uncertainty, for different survey masks, random mock data seeds, bias parameters, and resolutions (reso.). The main result of this paper is run \#4.}
\label{tab:results}
\end{table}

\begin{figure}
	\center
	\includegraphics[width=1.0\columnwidth]{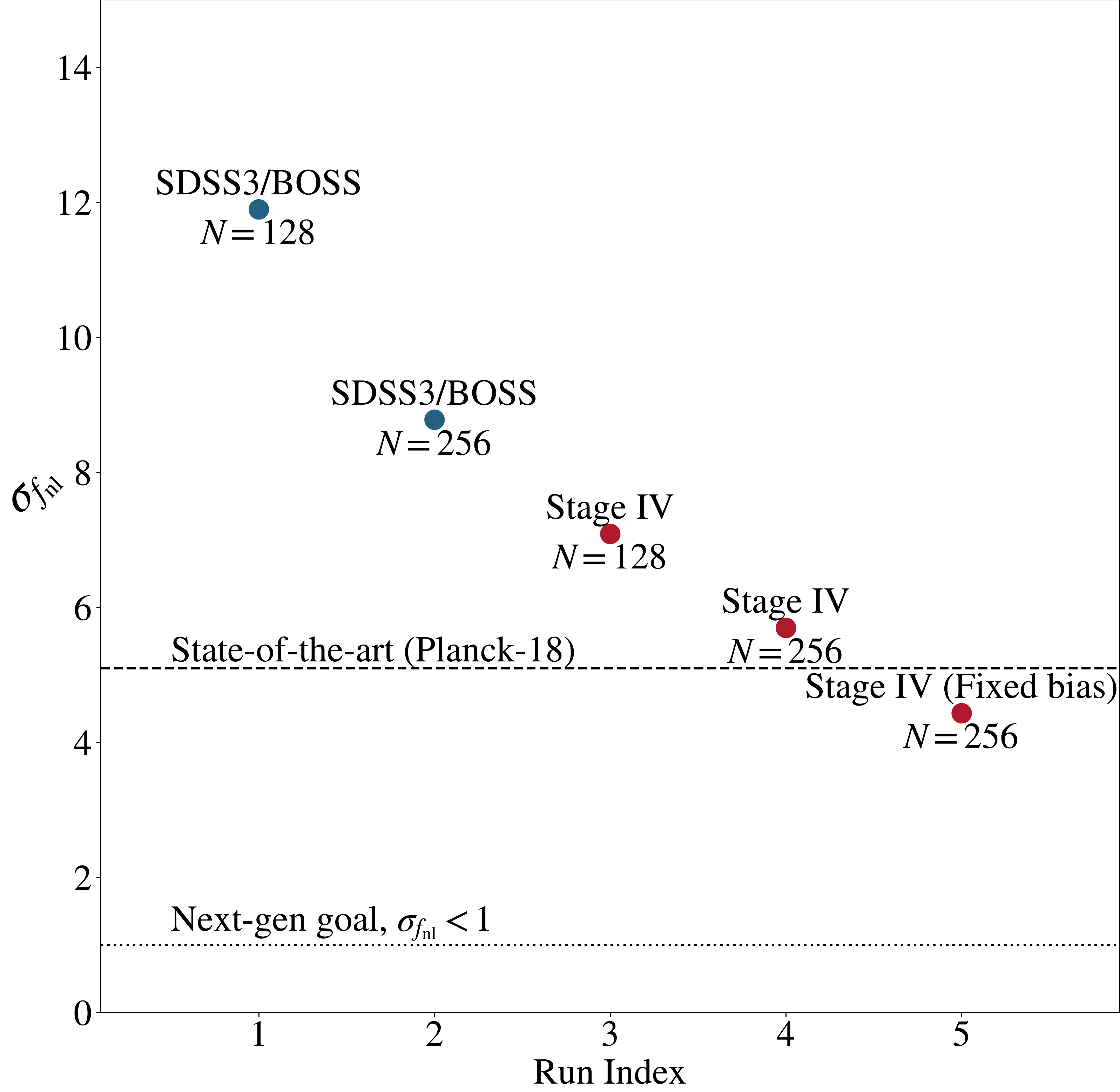}
	\caption{Summary of the results of the runs performed. The vertical axis illustrates the estimated $\fnl$ uncertainty for each test run in this paper, as indicated by the horizontal axis. The blue dots represent the \sdssiii{}-like runs, and the red dots represent the \textit{Stage IV} runs. Notice how the increase in resolution improves the error on $\fnl$, from left to right in each data set. For the last two points, notice the impact of running the analysis with fixed bias parameters compared to marginalising them out.}
	\label{fig:sigma_plot}
\end{figure}

\begin{figure*}
	\center
    \includegraphics[width=2.0\columnwidth]{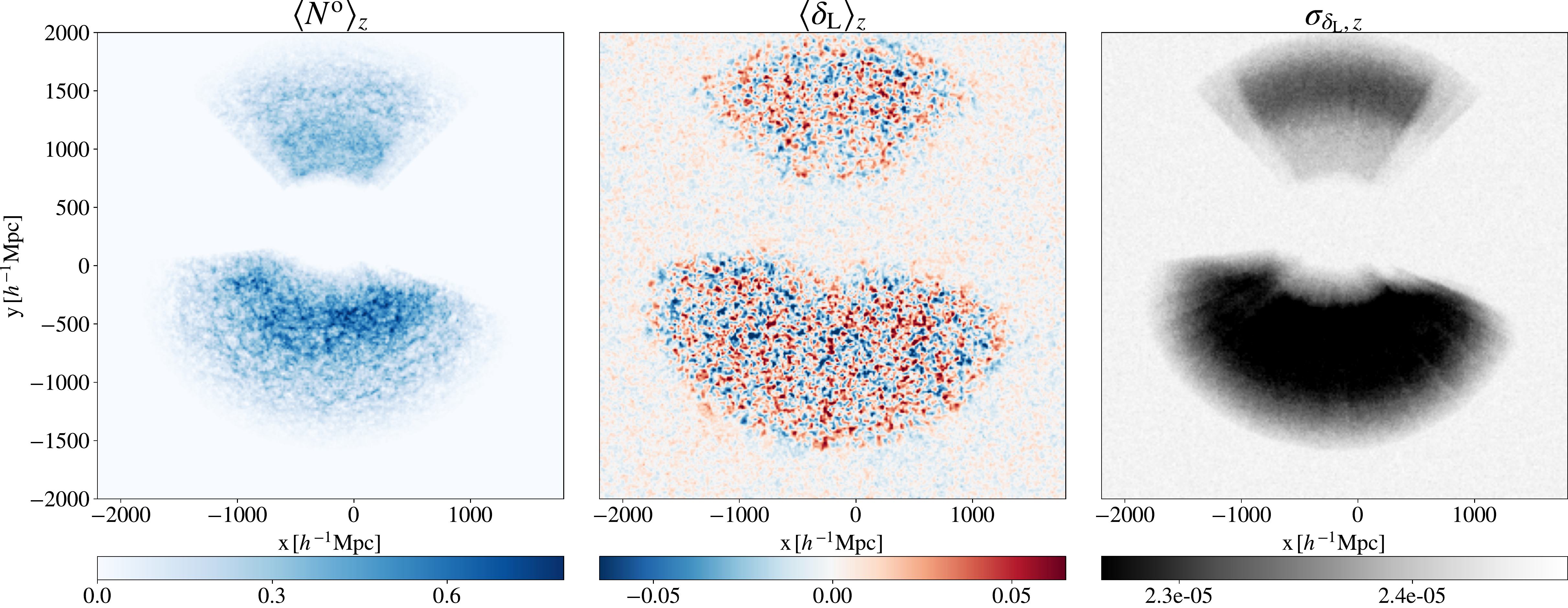}
	\caption{The galaxies of the \sdssiii{}-like run ($N=256$) (left panel), together with the mean (middle panel) and the standard deviation (right panel) over samples of the inferred linear density field, averaged over the $z$-axis. The galaxies represent the distribution of information in the data. In the mean linear density fields, regions containing galaxies are used by \borg{} to constrain the linear density field. In voxels absent of information, the density field is only constrained by the physical model and thus converges to the cosmic mean. In the standard deviation plot, the fluctuations are greatly reduced in galaxy-filled regions, as opposed to regions which lack observed data.}
	\label{fig:ensemble}
\end{figure*}

In Fig. \ref{fig:trace_stage_iv}, the $\fnl$-posterior distribution of run \#4 is plotted. Noticeably, the distribution displays little skewness, but is a near-Gaussian distribution, with the ground-truth $\fnl$-value within one $\sigma_{\fnl}$.

For each run listed in Table \ref{tab:runs}, we have computed the uncertainty on the sampled $\fnl$ value. These results are summarised in Fig. \ref{fig:sigma_plot}. As can be seen, \borg{} is able to constrain $\fnl$ to $\sigma_{\fnl} \approx 9$ for a \sdssiii{}-like data set, with a resolution at $256^3$. For a \textit{Stage IV} galaxy survey, \borg{} can bring down the error estimate to $\sigma_{\fnl} \approx 6$. In addition, we note the impact of the resolution: by increasing the resolution by a factor of $2$, we can bring down the error by a factor of $\sqrt{8}$, when changing the resolution from $128^3$ to $256^3$. 

For the interested reader, a full listing of all posterior distributions can be found in the appendix; see Section \ref{all_trace_plots}.

\subsection{Fixed bias test case}

In addition to performing resolution studies, we also performed a simplified test case, as outlined in Section~\ref{stage_iv}, representing the fixed bias scenario. Using high-resolution \textit{Stage IV} mock data, we sample the density field and $\fnl$-parameter, while keeping the bias parameters fixed. This is to test the scenario in which we have complete knowledge of the bias parameters. The improvement on the constrained $\fnl$-value can be seen in Table \ref{tab:results}. For a graphical visualisation, see Fig. \ref{fig:sigma_plot}. The outcome of the test shows that, for our simplified test case, having perfect information of the galaxy bias parameters can yield improvements by a factor of $0.78$ on $\sigma_{f_{\mathrm{nl}}}$, leading to $\sigma_{f_{\mathrm{nl}}} \approx 4.43$. 

\subsection{Accessing the field of primordial matter fluctuations}
\label{infer_dfield}

The Bayesian forward modelling approach goes beyond the estimation of statistical summaries, by fully incorporating the three-dimensional initial density field from which the observed structures formed. In this work, the \borg{} algorithm infers the primordial gravitation potential from galaxy mock data, which emulates the \sdssiii{} survey or the \textit{Stage IV} surveys. As a result, \borg{}'s Markov chain provides a numerical representation of the posterior distribution of three-dimensional initial density fields. Specifically, the Markov chain produces an ensemble of about $\approx 50000$ initial density field realisations that are conditioned on observations made at the present epoch. Given this ensemble, we can estimate any desired statistical summary of the Markov chain. For illustration, we plot the estimated mean and ensemble variance fields of the primordial initial conditions in Fig. \ref{fig:ensemble}. Notice how informative regions, i.e., data-filled regions, in the galaxy plot (left) correspond to regions where we have a stronger signal in the mean linear density field, $\langle \delta_L \rangle$, (middle) and less variance (right). In other words, the observed regions allow \borg{} to constrain the underlying $\delta_L$-field, while in masked regions, the $\delta$-field is only constrained by prior coupled to the physical model.

\begin{figure*}
\centering
\includegraphics[width=2.0\columnwidth]{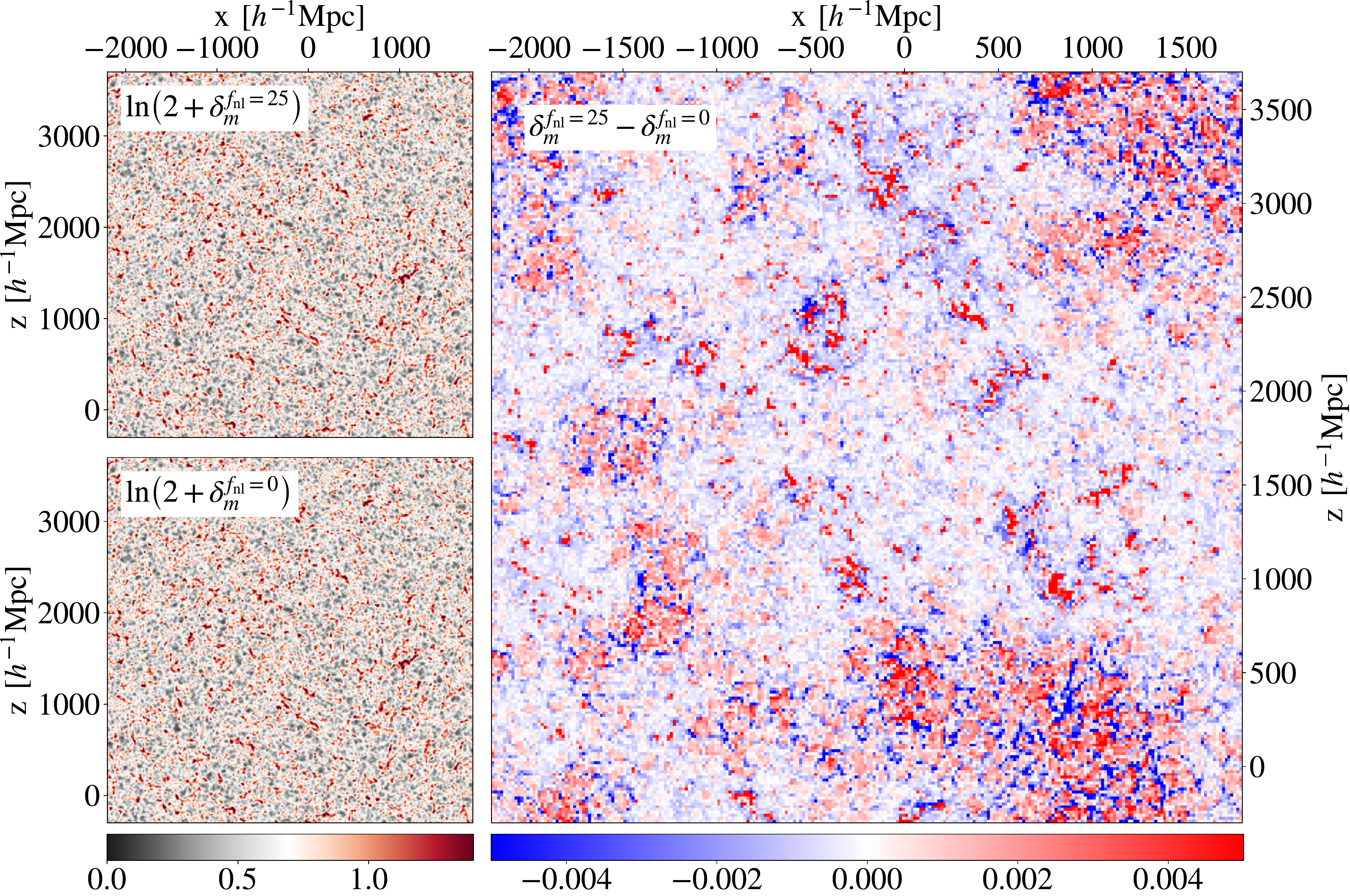}
\caption{The two left panels each display two identical fields generated by the same seed, varying only the $\fnl$ value ($25$ and $0$). The difference in the same-seed fields are plotted in the right panel, which depicts the regions of increased and decreased matter as a response to the changing $\fnl$ value. These maps illustrate the difference maps as a function of $\fnl$, which is a novel data product that our method can infer.}
\label{fig:diffdens} 
\end{figure*}

By varying the $\fnl$ parameter for a given realisation of the present-day density field, one can illustrate the PNG imprints on the cosmic LSS. An example of such a map can be found in Fig. \ref{fig:diffdens}. This is generated by subtracting two density fields originating from the same linear density field, which have been forward modelled using two different values of $\fnl$ ($25$, and $0$). In this map, hot spots and cold spots represent positions in a universe where PNG has increased or decreased the amount of matter (in relation to a universe without PNG). In principle, these maps could be used by observational missions to further probe the signals of PNG, e.g., searching for small-scale effects of $\fnl$ or studying the properties of galaxies in these regions. In other words, our method can highlight regions where PNG is expected to be imprinted given the data \citep{kostic2021machinedriven}.

\subsection{Tests of the scale-dependent bias model}
\label{model_tests}

As a concluding discussion, we outline the internal correlations of the model parameters. This is visualised in terms of marginalised posteriors of parameters, and contour plots. For the high resolution mock data of the \sdssiii{} run, the cross-correlations can be found in Fig. \ref{fig:pyramid}. The figure displays the marginalised posteriors of $\fnl$ and bias parameters, and the corresponding cross-contour plots. The fiducial values of the mock data are marked with blue lines. Notice how we are able to recover unbiased results of bias parameters (as the fiducial value falls within the $1\sigma$ range of the sampled values). More importantly, we are able to sample $\fnl$ while marginalising the unknown bias parameters without any strong degeneracy within the physical model. This suggests that we can jointly and independently sample $\fnl$ together with the bias parameters, without introducing any degeneracies between the model parameters.

We also perform a variety of diagnostic tests; these have the purpose of, among others, investigating the state of the MCMC chains and the robustness of the galaxy bias model. These tests include the correlation matrix and correlation lengths of model parameters. The resulting plots, together with a longer discussion, can be found in the Appendix (Section \ref{additional_results}).

\section{Summary and Conclusion}
\label{conclusions}

This work presents a novel approach to investigating the physics of the primordial universe, by using Bayesian physical forward modelling of the three-dimensional galaxy distribution in surveys. In contrast to traditional approaches (using only a limited set of statistical summaries), our method provides a full characterisation of three-dimensional density and velocity fields including relevant observable signatures of PNG. In addition, our method can measure galaxy bias parameters and provide estimates on their uncertainties. Specifically, our model directly infers information on the cosmic initial conditions and PNG by fitting a physical structure formation model to observations. By modelling data at the field level, our Bayesian forward modelling approach permits us to naturally and fully self-consistently exploit the full phenomenology that PNG imprints on the cosmic large-scale structure, e.g. cluster and void abundances, velocity fields, higher-order statistics, and scale-dependent galaxy bias. Moreover, \borg can simultaneously explore all relevant effects (e.g., both the primordial and structure-growth contributions to the bispectrum), and in this way break parameter degeneracies.

\begin{figure*}
	\center
	\includegraphics[width=2.0\columnwidth]{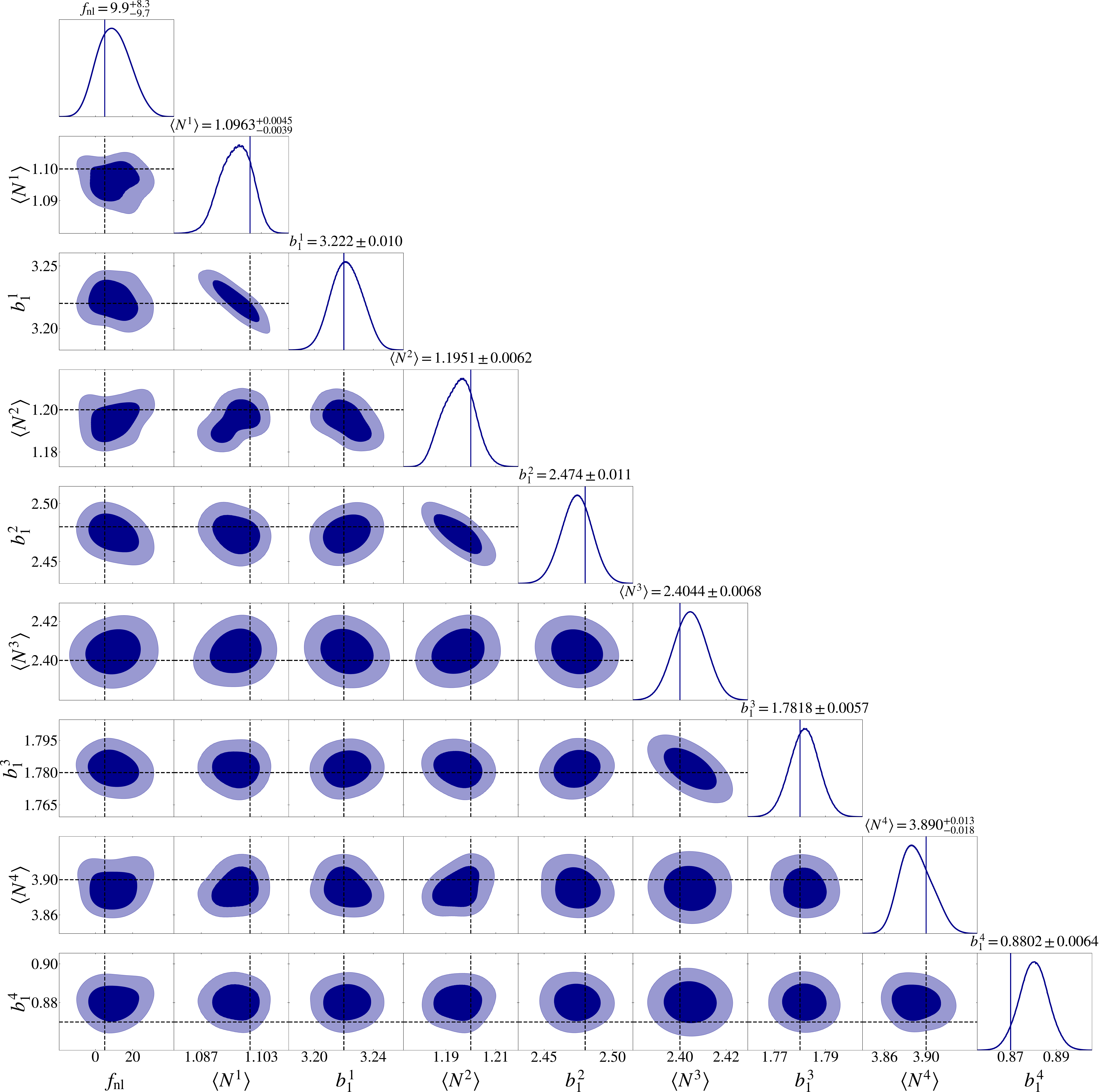}
	\caption{Cross-correlations and marginal posteriors for all hyperparameters and $\fnl$, for the high-resolution \sdssiii{}-like run. The darker regions in the contour plots represent the $1\sigma$ interval, the lighter regions represent the $2\sigma$ interval. The blue lines mark the fiducial values injected into the mock data, and constitute the true values of the data. The results indicate that our method is able to consistently sample the correct hyperparameters jointly with the true $\fnl$ parameter, since the lines overlap well with the contour plots. This demonstrates that our method can jointly infer the bias parameters of the galaxy bias model, and marginalise these out to constrain the $\fnl$ parameter.}
	\label{fig:pyramid}
\end{figure*}

To achieve this goal, we have expanded the Bayesian inference algorithm \borg to account for physical models of primordial non-Gaussianity in the generation of initial conditions. Here, we illustrated our method in the case of PNG of the local type, as parameterised by the cosmological parameter $\fnl$.
As discussed in Section~\ref{Method}, the inclusion of a physical model of PNG to \borg's forward model enters at two points in the large-scale structure posterior distribution. Firstly, PNG affects the generation of the cosmic initial conditions with the amplitude of non-Gaussian contributions scaled by $\fnl$. When evaluated with \borg's structure growth model, non-Gaussian initial conditions will directly yield imprints on the nonlinear matter and velocity distributions.

Secondly, PNG affects the formation of galaxies, which we account for by using a model of scale-dependent galaxy bias in our data model. The updated physics model therefore naturally and fully self-consistently accounts for a large phenomenology of PNG that is traditionally analysed independently. Besides, our field-level forward modelling approach provides direct solutions to account for selection effects and survey geometry, including measuring galaxy bias parameters.
At this point it should be remarked that our physics-informed inference framework can also handle even unknown survey and foreground contamination, a feature that will be particularly relevant for the inference of PNG in the next-generation surveys \citep[][]{2019A&A...624A.115P}. A final note: since the perturbation of $\fnl$ and gravitational evolution occur in two different steps in the forward model, \borg{} can differentiate between these two sources of nonlinearity in the data analysis procedure.

To test and illustrate the performance of the method, we applied it to two artificial galaxy surveys emulating the main survey characteristics of the \sdssiii{} and the upcoming \textit{Stage IV} surveys (see Section \ref{mock_observations}). The tests accounted for the respective survey geometries, selection functions, and noise. We chose to infer the corresponding initial conditions on a Cartesian equidistant grid of side length $L=4000\Mpch$ and $L=8000\Mpch$. We also performed a resolution study by performing inferences with $N_{\text{grid}}=128$ and $N_{\text{grid}}=256$, resulting in grid resolutions ranging from $\Delta L=\sim15.6\Mpch$ to $\Delta L=\sim62.5\Mpch$. We also configure a run with fixed bias. This indicates that \borg is able to handle unknown galaxy biases in a robust framework. Given this coarse resolution, we chose to approximate the gravitational structure formation using a Lagrangian perturbation theory model. The investigation of higher resolution with \borg's particle mesh simulation will be the subject of a coming publication. 

The outcomes of these tests have been presented and discussed in Section \ref{results}. To investigate the Markov chain's initial warm-up phase, we monitored the sequence of inferred posterior matter power spectra. The test revealed that the Markov chain completed the expected coherent drift towards the relevant region of the parameter space that contains most of the probability weight. From these tests we conclude that the initial warm-up phase takes about $7000$ Markov transitions, after which we start recording samples for the analysis. The test further reveals that the Markov chain correctly accounts for the survey geometries and selection effects by inferring the correct shape of the cosmic power spectrum throughout all Fourier modes considered in this analyses. Note that the correct inference of large-scale power is of importance for the inference of PNG in general and $\fnl$ in particular.

One of the major outcomes of this work is the demonstration that the $\fnl$ parameter can be constrained with a field-level forward modelling approach. We demonstrate that \borg{} can constrain $\fnl$ to $\sigma_{\fnl}=8.78$ for a \sdssiii{}-like survey, and $\sigma_{\fnl}=5.7$ for a \textit{Stage IV} survey. A comparison with other papers has been summarised in Table $\ref{tab:comparison}$. We studied the inference at two different resolutions, showing that the extracted information on $\fnl$ increases when using higher resolution. This result corresponds to our expectations: we expect the number of usable modes for the analysis to scale as $k^3$. The results suggest that more information can be extracted by increasing the resolution further, which will be investigated in upcoming studies. Another key point is the fact that our method automatically marginalises out the nuisance parameters associated with the galaxy bias model. This is an essential feature of the model, since it allows for unbiased estimates of cosmological parameters. As illustrated by Fig. \ref{fig:pyramid}, \borg{}'s Markov chain Monte Carlo accurately accounts for and corrects the correlations between the parameter of interest, $\fnl$, and other nuisance parameters. When having complete knowledge of the galaxy bias parameter for a \textit{Stage IV} data analysis, we find that our method can constrain $\fnl$ to $\sigma_{\fnl}=4.43$. For more tests and their results, the interested reader is referred to the appendix (Section \ref{additional_results}). In other words, by using all higher order statistics of galaxy clustering, we can utilise more of the data than traditional data analysis tests \citep{McQuinn_2021}.

A major advantage of our approach is its ability to characterise the full three-dimensional field of initial density fluctuations. In Section \ref{infer_dfield}, we show inferred three-dimensional initial conditions. In contrast to traditional methods that either report only cosmological parameters or compressed statistical summaries, our approach therefore provides the possibility of performing posterior predictive tests on complementary data sets. Besides opening the possibility to search for new physics via cross-correlation with complementary data, the possibility of performing detailed posterior predictive tests adds to the reliability and robustness of our approach. For instance, this opens the immediate opportunity to confirm or detect PNG by cross-correlating our inferences with next-generation CMB data to study the kinetic Synyaev-Zeldovich effect \citep[see e.g.][]{2020JCAP...12..011N}. One can, in principle, fold in the kinetic Synyaev-Zeldovich observable to jointly constrain $\fnl$ in the same field-level forward modelling framework, e.g., \borg{}. 

In summary, this work presents the first fully Bayesian forward modelling solution to investigate primordial universe physics with galaxy surveys. The results demonstrate the potential of jointly exploiting the entire phenomenology that PNG imprints on the three-dimensional density and velocity fields. Besides providing estimates of the $\fnl$ parameter, the method produces maps of three-dimensional density and velocity fields permitting posterior predictive cross-correlation studies with complementary data. By accurately accounting for major survey systematic effects and marginalising out bias parameters, our method can infer $\fnl$ at the level of $5.70$ on next-generation galaxy surveys. Future work will be focused on increasing the resolution of the reconstructions, thereby resolving more information for constraining PNG. Our results demonstrates the promise of Bayesian forward modelling to study the physics of the origin of the Universe with next-generation galaxy surveys.

\begin{table}[]
\centering
\begin{tabular}{lll}
\hline
Author           & $\sigma_{\fnl}$        & Survey      \\ \hline
This paper   & $9$ & \sdssiii{} (mock)  \\ 
\cite{mueller2021clustering}  & $21$     & SDSS-IV/eBOSS \\
\cite{damico_limits_2022} & $29$    & \sdssiii{} \\ \hline
\end{tabular}
\caption{Comparison of our inferred $\fnl$ constraints with other papers measuring $\fnl$ in galaxy redshift surveys. Note, while the two last data sets are observed with the same instrument, they cover different regions in the Universe and have detected different tracers. }
\label{tab:comparison}
\end{table}

\section*{Acknowledgements}
We are grateful to many enlightening discussions with Steffen Hagstotz, Alexandre Barreira, Fabio Finelli, and Yashar Akrami. We would also like to thank Eleni Tsaprazi, Nhat-Minh Nguyen and Natalia Porqueres for providing feedback on the manuscript. AA acknowledges the travel funding supplied by the University of Oxford, and by the Birger och Gurli Grundströms forskarstipendiefond. JJ acknowledges support by the Swedish Research Council (VR) under the project 2020-05143 -- "Deciphering the Dynamics of Cosmic Structure". GL acknowledges support by the ANR BIG4 project, grant ANR-16-CE23-0002 of the French Agence Nationale de la Recherche.
FS acknowledges support from the Starting Grant (ERC-2015-STG 678652) ``GrInflaGal'' of the European Research Council. This work was supported by the Simons Collaboration on ``Learning the Universe''.

The computation and data processing in this study were enabled by resources provided by the Swedish National Infrastructure for Computing (SNIC) at Tetralith, partially funded by the Swedish Research Council through grant agreement no. 2020-05143. 
This research utilised the HPC facility supported by the Technical Division at the Department of Physics, Stockholm University.
This work is done within the Aquila Consortium\footnote{\url{https://www.aquila-consortium.org/}}. 
We acknowledge the use of the following packages: \texttt{NumPy} \citep[]{harris2020array}, \texttt{Matplotlib} \citep[]{Hunter_2007}, \texttt{GetDist} \citep[]{lewis2019getdist}, and \texttt{HEALPix} \citep[]{gorski_healpix_2005}.

\section*{Data Availability}

The data underlying this article will be shared on the basis of a reasonable request to the corresponding author.

\newpage

\bibliographystyle{mnras}
\bibliography{mnras}

\appendix

\section{Additional Results}
\label{additional_results}

\begin{figure*}
  \subcaptionbox{The marginalised posterior distribution of $\fnl$ for run \#1; the low resolution \sdssiii{}-like run. \label{fig:pdf_1}}{\includegraphics[width=1.65\columnwidth]{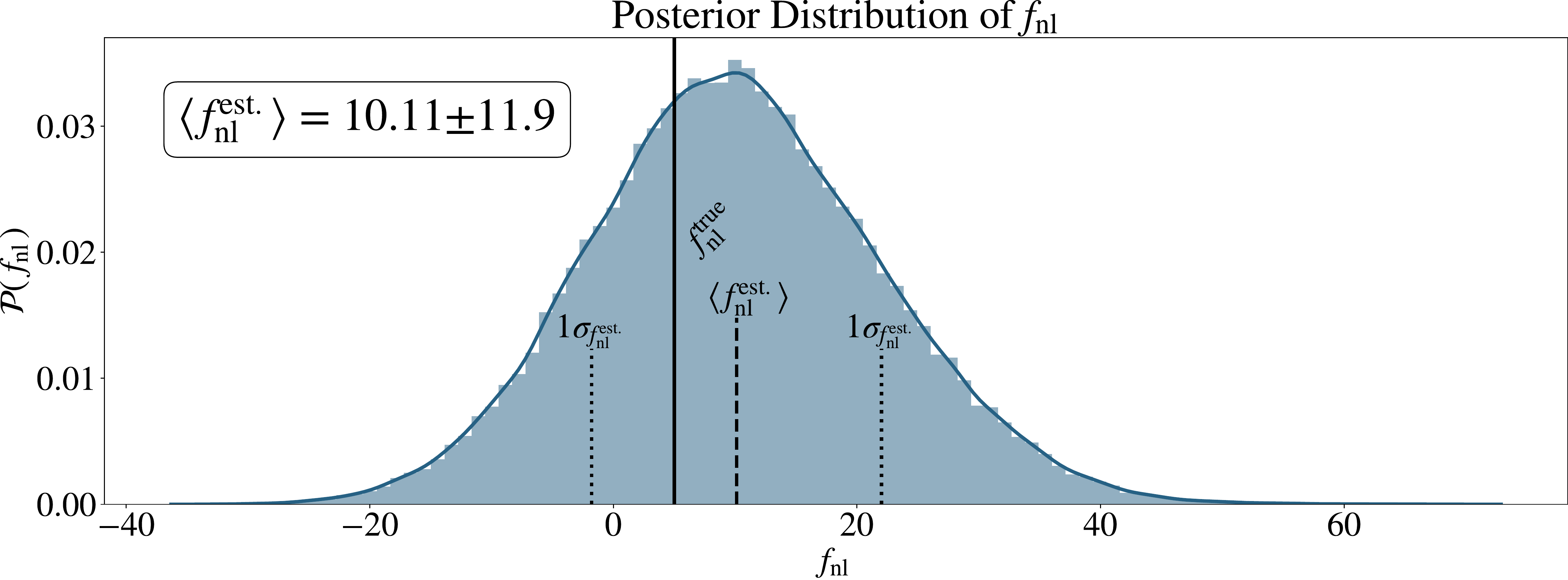}}
  \subcaptionbox{The marginalised posterior distribution of $\fnl$ for run \#2; the high resolution \sdssiii{}-like run. \label{fig:pdf_3}}{\includegraphics[width=1.65\columnwidth]{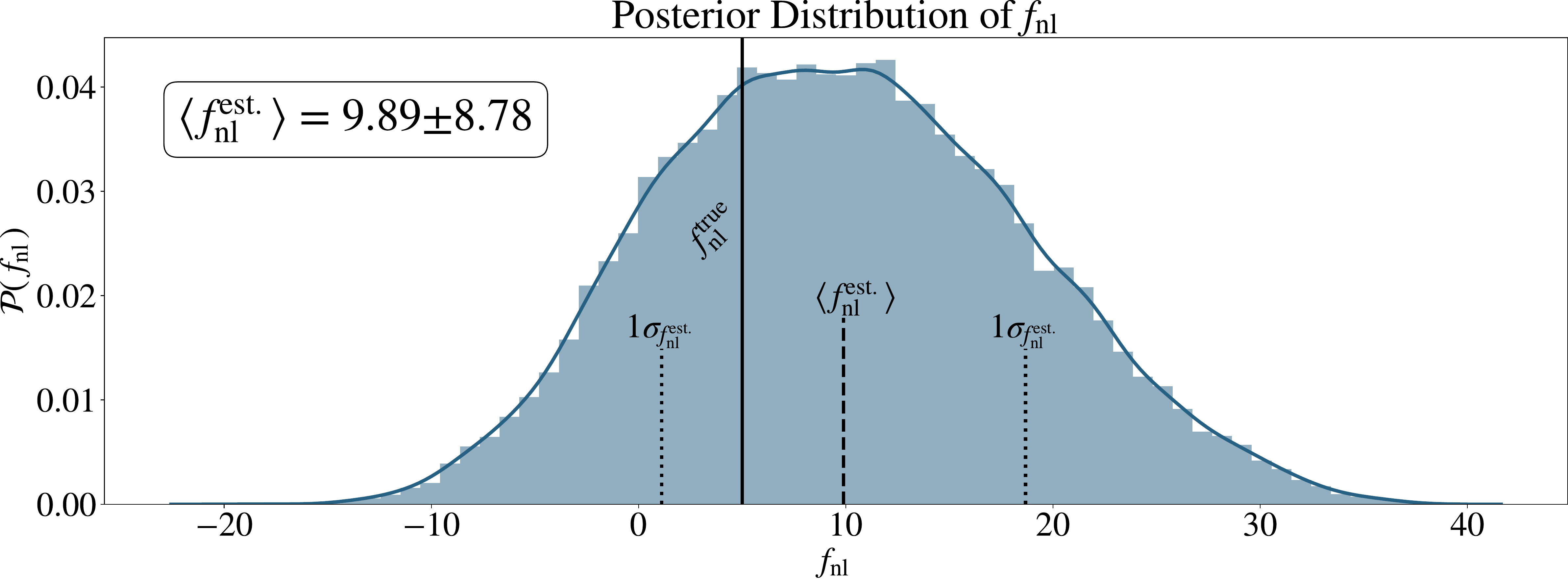}}
  \subcaptionbox{The marginalised posterior distribution of $\fnl$ for run \#3; the low resolution \textit{Stage IV} run. \label{fig:pdf_2}}{\includegraphics[width=1.65\columnwidth]{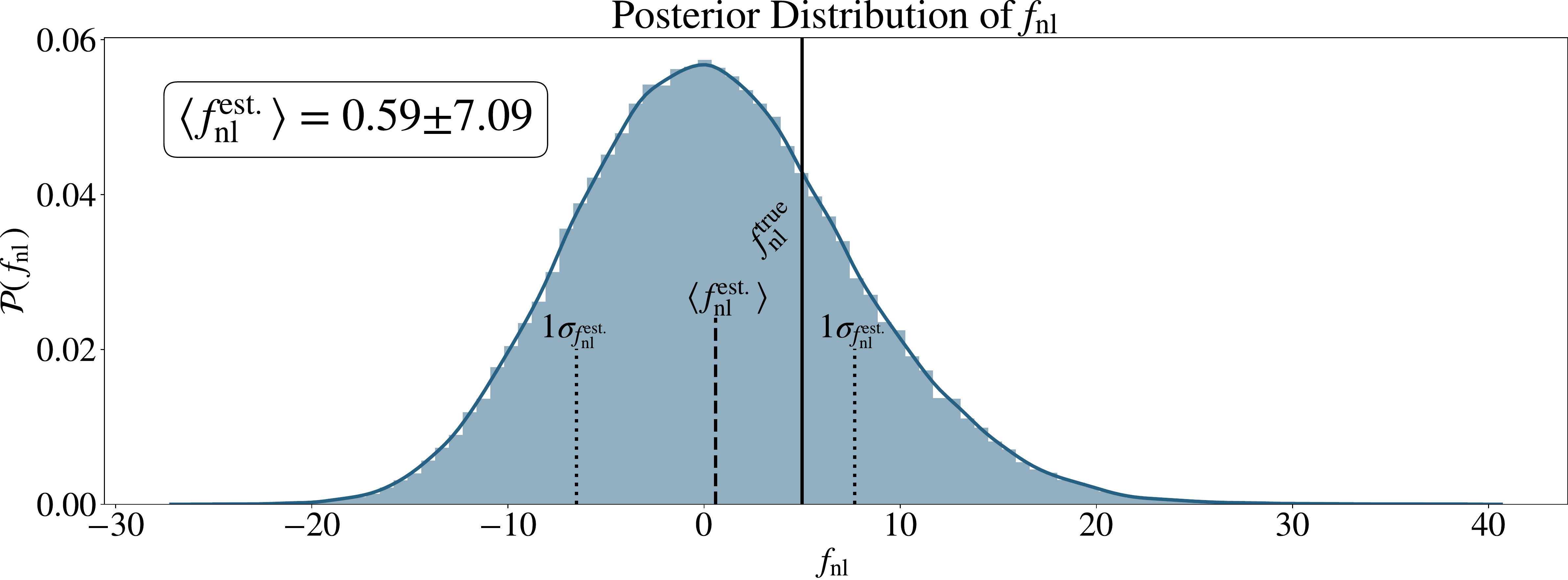}}
  \subcaptionbox{The marginalised posterior distribution of $\fnl$ for run \#5; the high resolution \textit{Stage IV} run, with fixed bias sampling. \label{fig:pdf_4}}{\includegraphics[width=1.65\columnwidth]{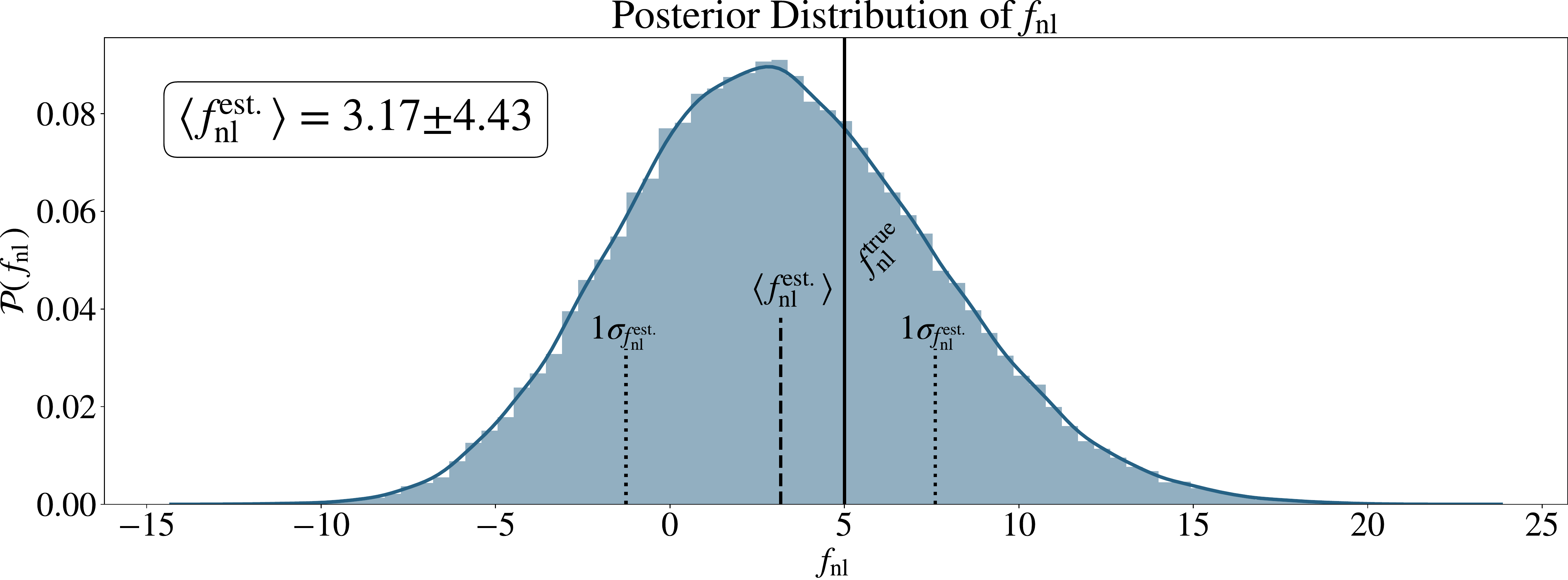}}
  \caption{The inferred posterior distributions of $\fnl$ for all other runs. These distributions include handling and marginalising out all other effects, e.g., survey geometry, instrumentation noise, and galaxy biases, with the exception of run \#5.}
  \label{fig:runs_pdfs} 
\end{figure*}

\begin{figure}
	\center
	\includegraphics[width=1.0\columnwidth]{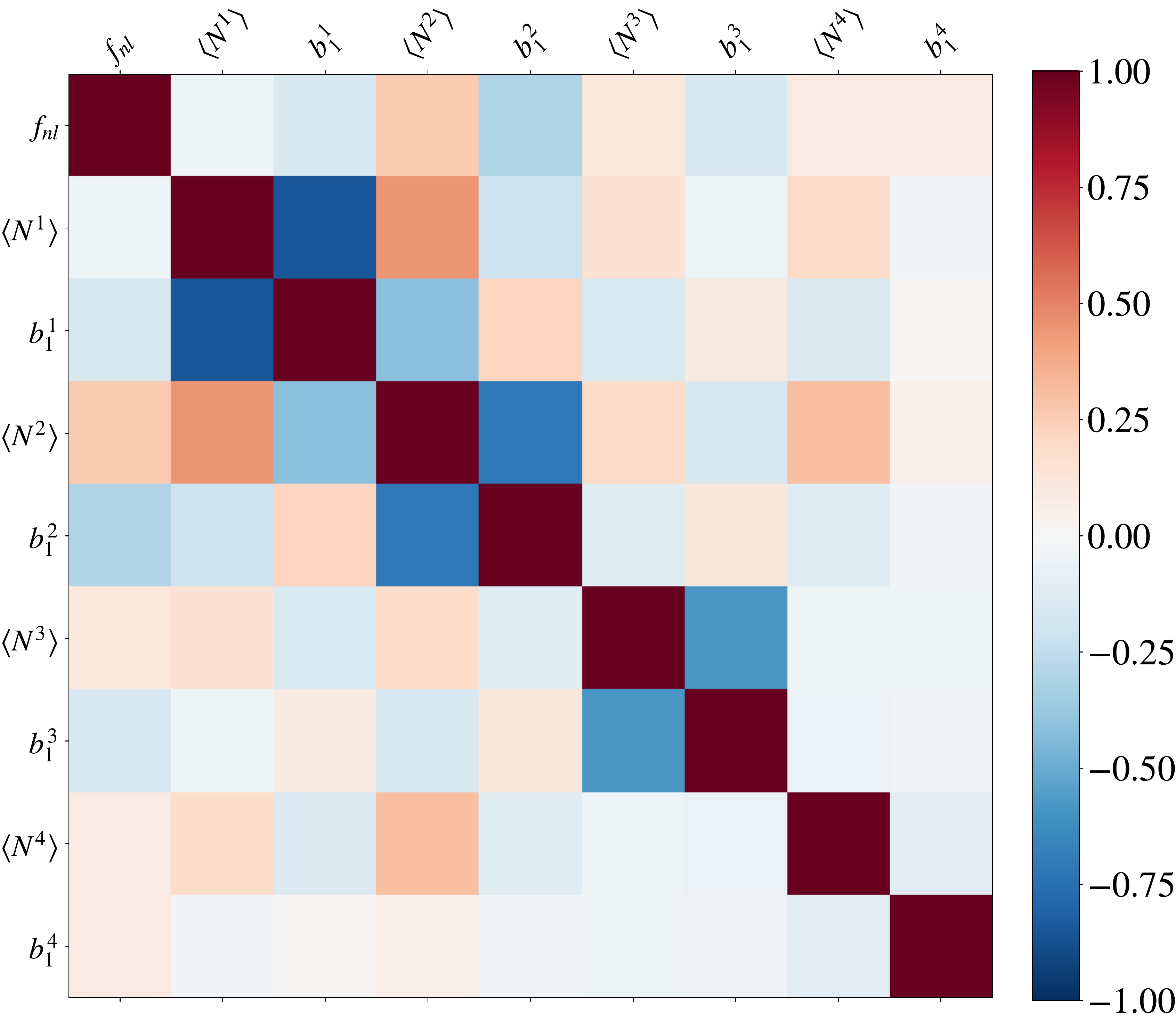}
	\caption{The plot shows the correlation matrix of run \#2. The correlation matrix displays any internal correlations between model parameters. Most importantly, any correlations between $\fnl$ and the galaxy bias parameters are undesired, since they would indicate that the inferred $\fnl$ may be biased by the model parameters. The absence of these features indicate that we can robustly infer the correct $\fnl$ value with $1\sigma$ confidence.}
	\label{fig:corr_mat}
\end{figure}

\subsection{$\fnl$ posterior distributions of runs}
\label{all_trace_plots}
In Fig. \ref{fig:runs_pdfs}, we have illustrated the posterior distributions of $\fnl$ for the secondary runs of this paper. While their uncertainties have been provided in Table \ref{tab:results} and in Fig. \ref{fig:sigma_plot}, here we provide the full inferred posterior distributions. Note that these distributions include any and all survey effects, e.g., survey geometry, instrumentation noise, and galaxy biases, except for run \#5. The posterior distributions have their centres within 1$\sigma$ of the fiducial values ($\fnl^{\mathrm{fid.}}=5$), which indicates that our method can consistently infer the correct $\fnl$-distribution.

\subsection{Correlation matrix}

The correlation matrix of the high resolution \textit{Stage IV} mock data run can be found in Fig. \ref{fig:corr_mat}. The figure displays the correlations and anti-correlations between model parameters throughout the samples in the chain. As can be seen, there is a very weak correlation between $\fnl$ and the bias parameters of the different tracer catalogues. This is a very promising fact since it indicates that there is little-to-no relationship directly between $\fnl$ and the hyper-parameters of the galaxy bias model. This implies that the cosmological parameter is being sampled and detected without strong dependencies on the bias parameters, which strengthens the validity of our method. Moreover, these results further suggest, together with the figure of contour plots (Fig. \ref{fig:pyramid}), that the model has very few degeneracies between the model parameters.

\subsection{Correlation lengths}

\begin{figure}
	\center
	\includegraphics[width=1.0\columnwidth]{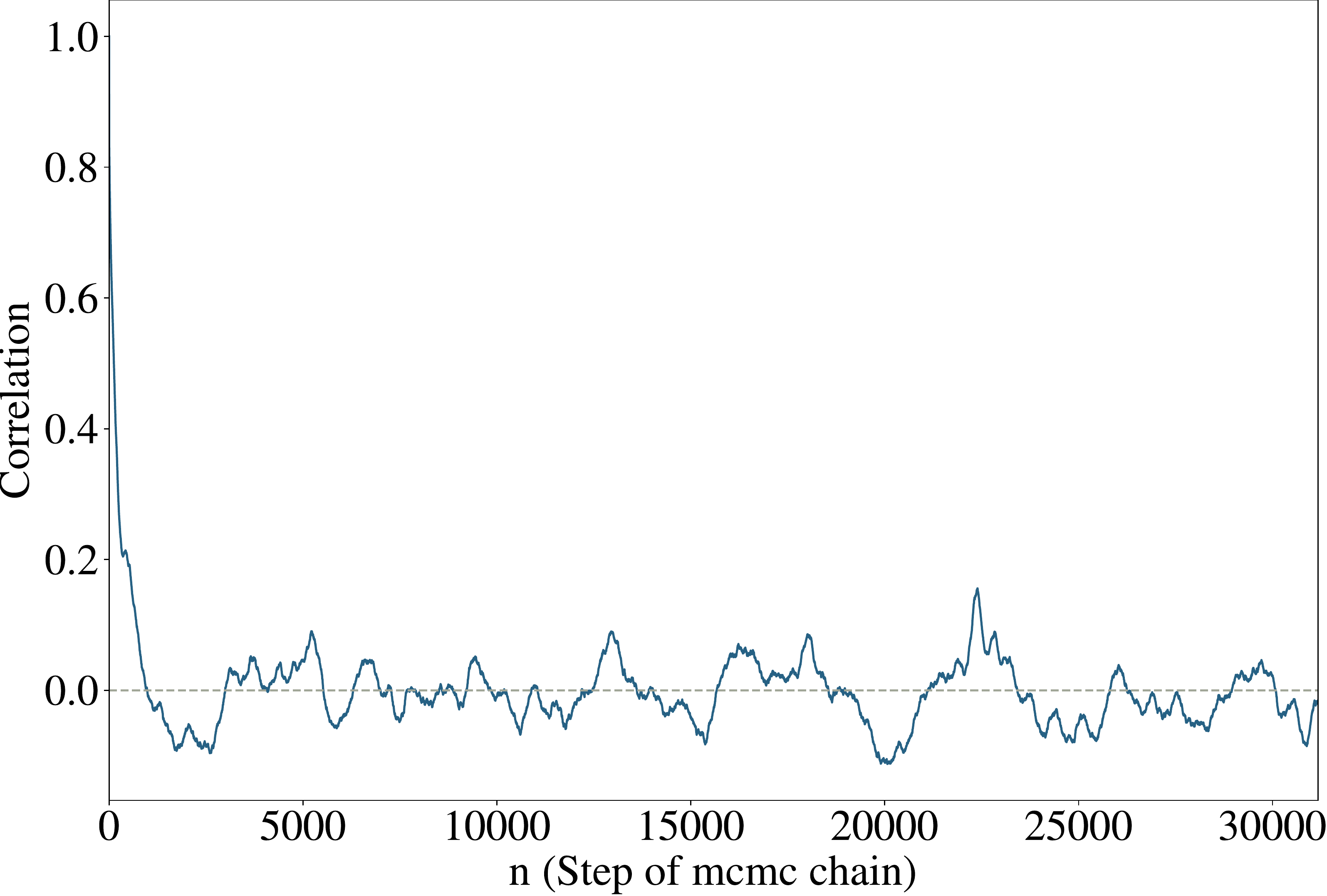}
	\caption{The plot shows the autocorrelation of sampled $\fnl$ values as a function of sample step of run \#4. Note in the figure how the samples become uncorrelated early in the chain (by dropping from 1 to 0 quickly), and oscillates around 0. This means that the number of steps needed for the MCMC algorithm to produce statistically independent samples are relatively few, compared to the overall length of the full chain.}
	\label{fig:corr_length}
\end{figure}

\begin{figure}
	\center
	\includegraphics[width=1.0\columnwidth]{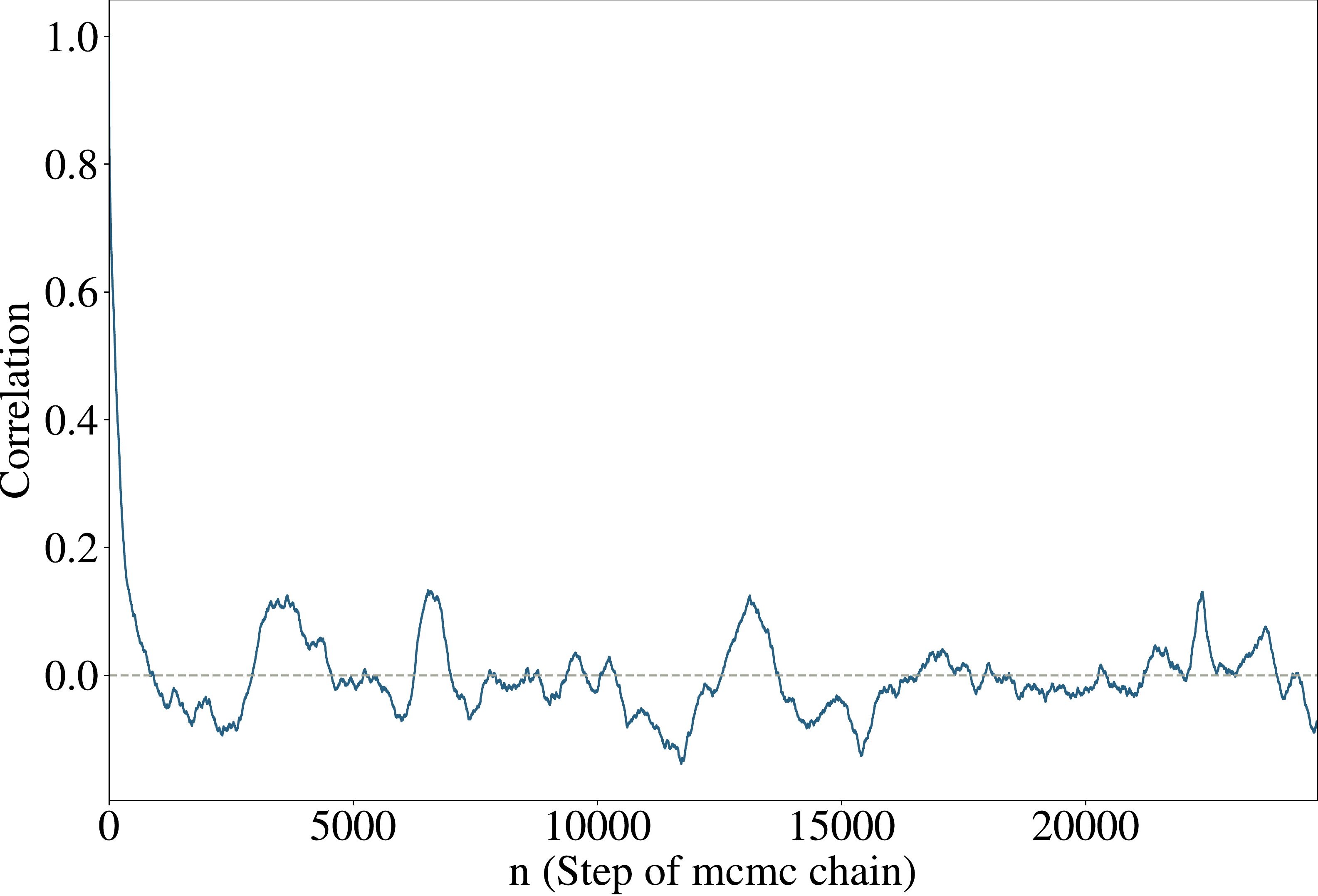}
	\caption{The plot shows the autocorrelation of sampled $\fnl$ values as a function of sample step of run \#5.}
	\label{fig:corr_length_fixed}
\end{figure}

The correlation lengths of the high resolution \textit{Stage IV} runs can be found in Fig. \ref{fig:corr_length_fixed} and \ref{fig:corr_length}. This plot displays the correlation of the $\fnl$ parameter between the samples of the chains. This can be described as: 

\begin{eqnarray}
    C(\fnl)_{\mathrm{n}} = \frac{1}{N-n}\sum^{\mathrm{N}-\mathrm{n}}_{\mathrm{i}=0} \left( \frac{f_{\mathrm{nl}}^\mathrm{i}-\langle\fnl\rangle}{\sigma_{\fnl}} \frac{f_{\mathrm{nl}}^{\mathrm{i}+\mathrm{n}}-\langle\fnl\rangle}{\sigma_{\fnl}}   \right ) \, ,
\end{eqnarray}
where $n$ is the number of transition steps, $\langle\fnl\rangle$ and $\sigma_{\fnl}$ are the mean and the standard deviation of the sampled $\fnl$ value, and $N$ is the total number of samples in the run. The plot shows that \borg{} is able to sample statistically independent samples roughly after $\approx 1000$ samples. These results demonstrate that the statistical efficiency of our $\fnl$ sampler is robust, since the number of steps between independent samples is much smaller than the total length of the chain.

\section{Survey configurations for the \textit{Stage IV} survey}
\label{stage_iv_mask}

\begin{figure}
	\center
    \includegraphics[width=1.0\columnwidth]{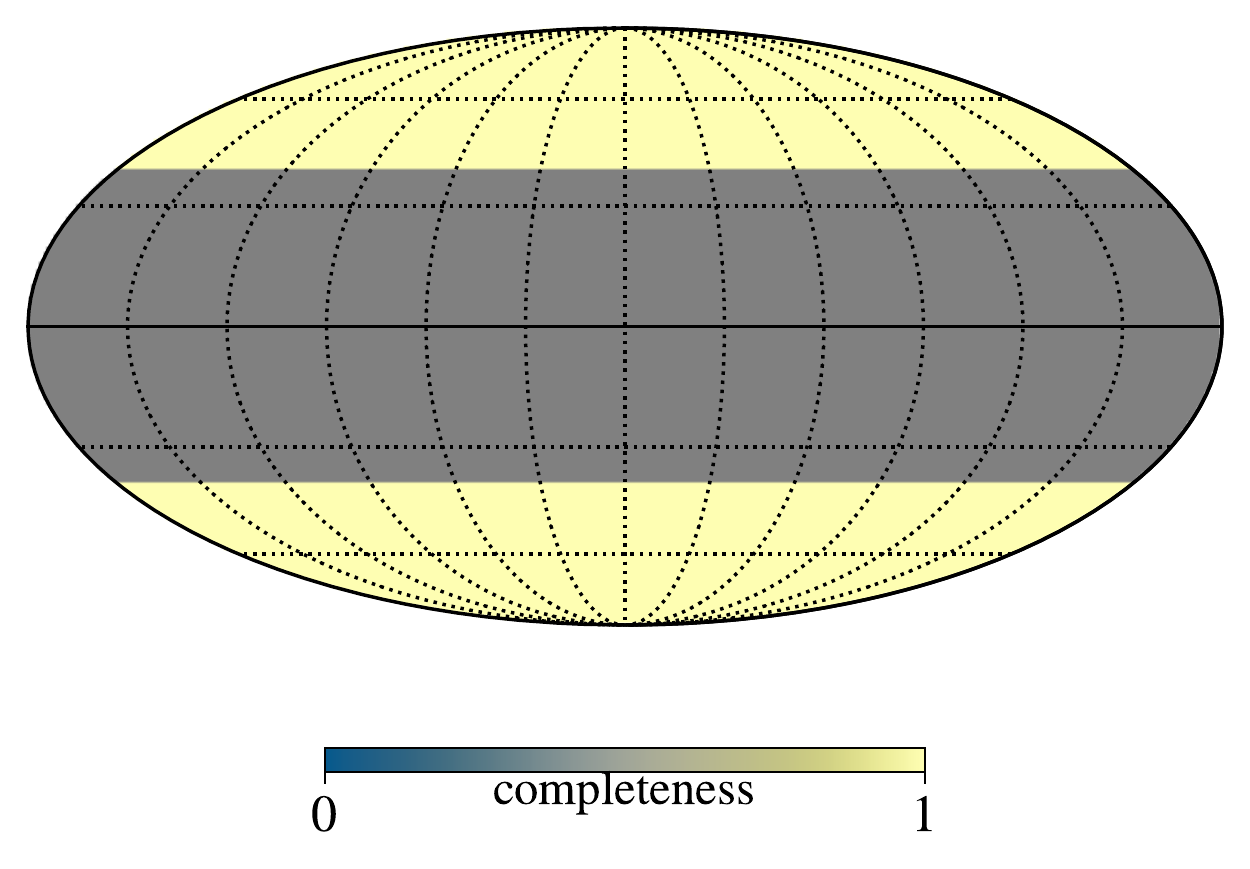}
	\caption{The sky map displaying the observed and masked regions for the \textit{Stage IV} runs of this study.}
	\label{fig:stage_iv_galaxy_masks}
\end{figure}

\begin{figure}
	\center
    \includegraphics[width=1.0\columnwidth]{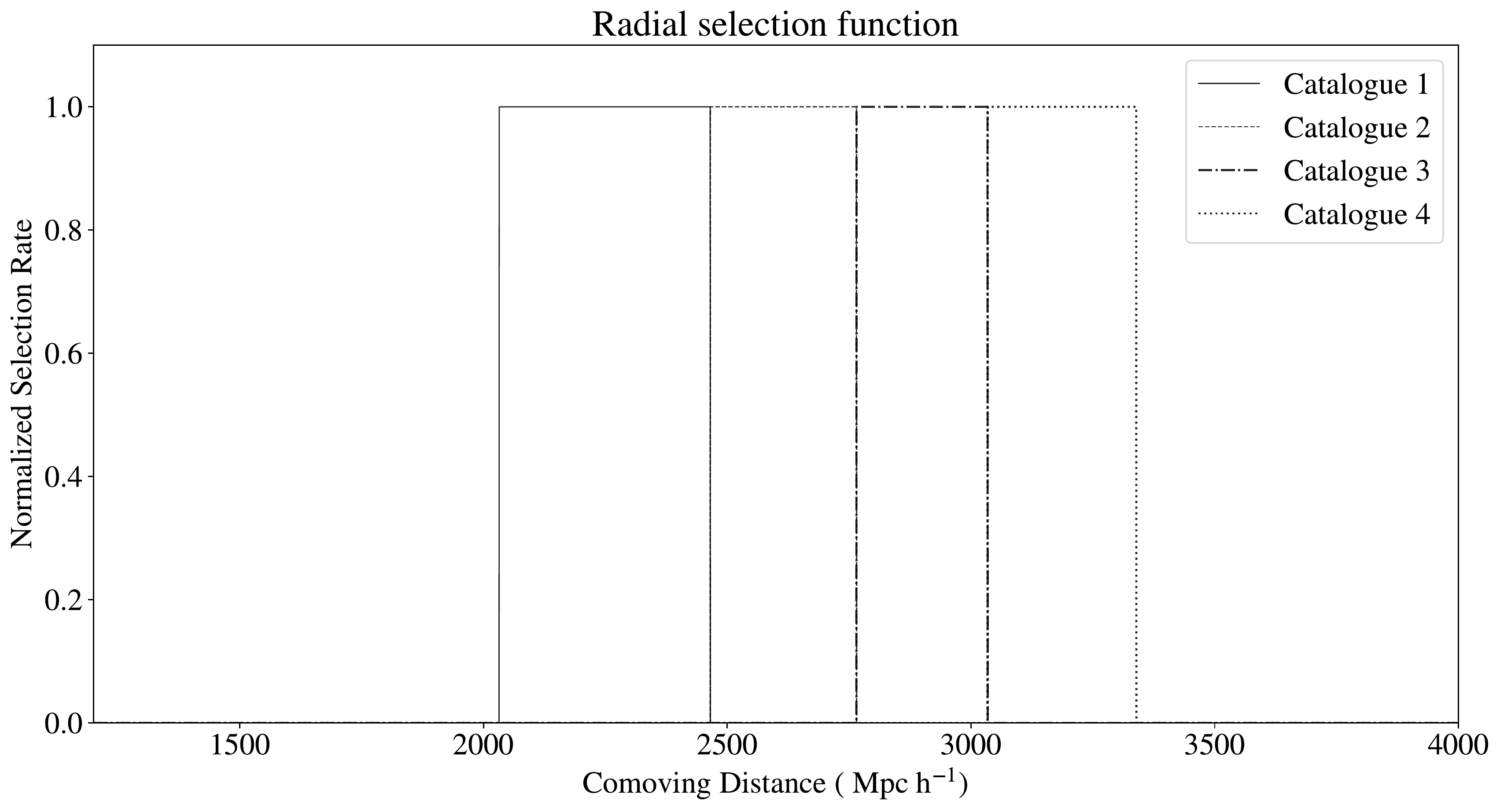}
	\caption{The radial selection functions for the \textit{Stage IV} runs of this study.}
	\label{fig:stage_iv_rs}
\end{figure}

In this section, we provide plots of the radial selection functions and the sky map for the \textit{Stage IV} survey. The design choices for these specifications are outlined in Section \ref{stage_iv}. The radial selection functions can be seen in Figure \ref{fig:stage_iv_rs}. The sky map can be seen in Figure \ref{fig:stage_iv_galaxy_masks}.

\section{Gelman-Rubin test}
\label{gr_test}
In this section, we outline the Gelman-Rubin test, which evaluates the convergence of a MCMC chain \citep{gr_test_1992}. The process in which it does so is the following: First, the chain is divided into $\mathrm{M}$ different subchains, each containing $\mathrm{n}$ samples, with $\mathrm{d}$ in-between (which are ignored in the analysis). Then, we compute the potential scale reduction factor:

\begin{eqnarray}
\mathrm{PSRF} = \sqrt{\frac{V}{W}}\, ,
\end{eqnarray}
where:
\begin{enumerate}
    \item $V = \frac{n-1}{n}W + \frac{M+1}{nM}$,
    \item $W = \frac{1}{M}\sum_{m=1}{M}\sigma_m^2$,
    \item $B = \frac{n}{M-1}\sum_{m=1}{M}(\langle \fnl \rangle_m - \langle \fnl \rangle)^2$.
\end{enumerate}
with $\sigma_m^2$ being the variance of $\fnl$ for the $\mathrm{m}$th chain. The chain is considered to be converged as $\mathrm{PSRF} \rightarrow 1$

For run \#4, we achieve the value of $V=1.000021$, when $M=2$, $n=25000$, and $d=6100$. Thus, this value indicates that the chain has converged \citep{gr_test_1992}. Similar findings are made for the other runs, as well.

\section{Adjoint Gradient Calculation of the $\fnl$ Perturbation}
\label{AG_fnl}

In this section, we outline the adjoint gradient of the $\fnl$ perturbation module, which is used by the algorithm to evaluate the sensitivity of model parameters. In this case, we want to evaluate the change in the non-Gaussian primordial density field $\delta^\mathrm{NG}$, as the input Gaussian density field $\delta^\mathrm{G}$ is altered. First, we drop the $m$-index for the matter density field $\delta_{\mathrm{m}}$, such that $\delta_{\mathrm{m}} \equiv \delta$, for the sake of brevity. Then, we start in the general case:

\begin{eqnarray}
\frac{\partial\mathrm{ln}(\pi(\delta^\mathrm{NG}))}{\partial\delta_\mathrm{j}^\mathrm{G}}
= \sum_{\mathrm{i}} \frac{\partial\mathrm{ln}(\pi(\delta^\mathrm{NG}))}{\partial\delta_{\mathrm{i}}^\mathrm{NG}}\frac{\partial\delta_{\mathrm{i}}^\mathrm{NG}}{\partial\delta_\mathrm{j}^\mathrm{G}} \, .
\end{eqnarray}

Next, we evaluate $\frac{\partial\delta_{\mathrm{i}}^\mathrm{NG}}{\partial\delta_\mathrm{j}^\mathrm{G}}$. But in order to do so, we must first decompose $\delta_{\mathrm{i}}^\mathrm{NG}$:

\begin{eqnarray}
\delta_{\mathrm{i}}^\mathrm{NG} = \sum_\mathrm{a} \mathcal{\bar{F}}_{\mathrm{i},\mathrm{a}}\Phi_\mathrm{a} \, , \nonumber \\
\Phi_\mathrm{a} = \phi_\mathrm{a} + \fnl\phi_\mathrm{a}^2 \, , \\
\phi_\mathrm{a} = \sum_\mathrm{b}\mathcal{F}_{\mathrm{a},\mathrm{b}}\delta^{\textrm{G}}_\mathrm{b} \, , \nonumber
\end{eqnarray}
where $\mathcal{F}$ is the Fourier transform, $\Phi$ is the perturbed gravitational potential and $\phi$ is the non-perturbed gravitational potential, both potentials transformed to redshift $z=1000$, and acted under the correct transfer functions. 

Now, we can evaluate $\frac{\partial\delta_{\mathrm{i}}^\mathrm{NG}}{\partial\delta_\mathrm{j}^\mathrm{G}}$:

\begin{eqnarray}
\frac{\partial\delta_{\mathrm{i}}^\mathrm{NG}}{\partial\delta_\mathrm{j}^\mathrm{G}} = \frac{\partial\delta_{\mathrm{i}}^\mathrm{NG}}{\partial\Phi_\mathrm{k}}\frac{\partial\Phi_\mathrm{k}}{\partial\phi_\mathrm{l}}\frac{\partial\phi_\mathrm{l}}{\partial\delta_\mathrm{j}^\mathrm{G}} \, .
\end{eqnarray}

The expressions for these partial derivatives are:

\begin{eqnarray}
\frac{\partial\delta_{\mathrm{i}}^\mathrm{NG}}{\partial\Phi_\mathrm{k}} = \sum_\mathrm{b}\mathcal{\bar{F}}_{\mathrm{i},\mathrm{b}} \delta^\mathrm{K}_{\mathrm{b},\mathrm{k}} \, , \nonumber \\
\frac{\partial\Phi_\mathrm{k}}{\partial\phi_\mathrm{l}} = \delta^\mathrm{K}_{\mathrm{k},\mathrm{l}} + 2\delta^\mathrm{K}_{\mathrm{k},\mathrm{l}}\fnl\phi_\mathrm{k} \, , \\
\frac{\partial\phi_\mathrm{l}}{\partial\delta_\mathrm{j}^\mathrm{G}} = \sum_\mathrm{q}\mathcal{F}_{\mathrm{l},\mathrm{q}}\delta^\mathrm{K}_{\mathrm{l},\mathrm{j}} \, , \nonumber
\end{eqnarray}

where $\delta^\mathrm{K}$ is the Kronecker delta. 

Now, combining yields the full expression of the adjoint gradient. The log-posterior distribution is therefore:

\begin{eqnarray}
\frac{\partial\mathrm{ln}(\pi(\delta^\mathrm{NG}))}{\partial\delta_\mathrm{j}^\mathrm{G}} = \sum_{\mathrm{i}}\frac{\partial\mathrm{ln}(\pi(\delta^\mathrm{NG}))}{\partial\delta_{\mathrm{i}}^\mathrm{NG}}\frac{\partial\delta_{\mathrm{i}}^\mathrm{NG}}{\partial\delta_\mathrm{j}^\mathrm{G}} &=& \nonumber \\
\sum_{\mathrm{i}}\frac{\partial\mathrm{ln}(\pi(\delta^\mathrm{NG}))}{\partial\delta_{\mathrm{i}}^\mathrm{NG}}\frac{\partial\delta_{\mathrm{i}}^\mathrm{NG}}{\partial\Phi_\mathrm{k}}\frac{\partial\Phi_\mathrm{k}}{\partial\phi_\mathrm{l}}\frac{\partial\phi_\mathrm{l}}{\partial\delta_\mathrm{j}^\mathrm{G}} &=& \nonumber \\
\sum_{\mathrm{i}}\frac{\partial\mathrm{ln}(\pi(\delta^\mathrm{NG}))}{\partial\delta_{\mathrm{i}}^\mathrm{NG}}[\sum_\mathrm{b}\mathcal{\bar{F}}_{\mathrm{i},\mathrm{b}} \delta^\mathrm{K}_{\mathrm{b},\mathrm{k}}][\delta_{\mathrm{k},\mathrm{l}} + \delta^\mathrm{K}_{\mathrm{k},\mathrm{l}}\fnl\phi_\mathrm{k}][\sum_\mathrm{q}\mathcal{F}_{\mathrm{l},\mathrm{q}}\delta^\mathrm{K}_{\mathrm{l},\mathrm{j}}] \, .
\end{eqnarray}

\section{Adjoint Gradient Computation of Bias Model}
\label{AG_sdb}

As in the previous section, we explicitly describe the adjoint gradient of the bias model. First, we drop the $m$-index for the matter density field $\delta_{\mathrm{m}}$, such that $\delta_{\mathrm{m}} \equiv \delta$, for the sake of brevity. Then, we perform the derivative of the log likelihood with respect to $\delta$:
\begin{eqnarray}
\frac{\partial \mathrm{ln}(\pi(\delta))}{\partial \delta_\mathrm{q}} = \sum_{\mathrm{i}} \frac{\partial \mathrm{ln}(\pi(\delta))}{\partial \rho_{\mathrm{i}}} \frac{\partial\rho_{\mathrm{i}}}{\partial\delta_\mathrm{q}} \, .
 \label{eq:diff_logLH}
\end{eqnarray}

Now we examine and expand the last term:

\begin{eqnarray}
\frac{\partial\rho_{\mathrm{i}}}{\partial\delta_\mathrm{q}} &=& \bar{n}_{\mathrm{gal}} \left ( b_1 + \frac{\partial \delta'_{\mathrm{i}}}{\partial \delta_{\mathrm{q}}} \right ),
\end{eqnarray}
where:

\begin{eqnarray}
\delta'_{\mathrm{i}}(\mvec{k},\fnl) = \Delta b(\mvec{k},\fnl) \delta_{\mathrm{i}}(\mvec{k}),
\end{eqnarray}
with $\Delta b(\mvec{k},\fnl)$ defined as in equation \ref{eq:b_ng}. Writing down each computation step, we get:

\begin{eqnarray}
\delta'_{\mathrm{p}}(\mvec{x}) = \sum_\mathrm{b}\mathcal{F}_{\mathrm{p},\mathrm{b}}\delta_\mathrm{b}(\mvec{k}) \, , \nonumber \\
\delta'_{\mathrm{j}}(\mvec{k}) = \Delta b(\mvec{k},\fnl) \delta_{\mathrm{j}}(\mvec{k})    \, , \\
\delta_{\mathrm{i}}(\mvec{k}) =  \sum_\mathrm{a} \mathcal{\bar{F}}_{\mathrm{i},\mathrm{a}}\delta_\mathrm{a}(\mvec{x}) \, , \nonumber 
\end{eqnarray}

Taking the partial derivatives with respect to the inputs (in the forward case), yields:

\begin{eqnarray}
\frac{\partial \delta'_{\mathrm{p}}(\mvec{x}) }{\partial \delta'_{\mathrm{j}}(\mvec{k}) } = \sum_\mathrm{b}\mathcal{F}_{\mathrm{p},\mathrm{b}}\delta^\mathrm{K}_{\mathrm{b},\mathrm{j}}
\, , \nonumber \\
\frac{\partial \delta'_{\mathrm{j}}(\mvec{k}) }{\partial \delta_{\mathrm{i}}(\mvec{k}) } = \Delta b(\mvec{k},\fnl) \delta^\mathrm{K}_{\mathrm{i},\mathrm{j}} \, , \\
\frac{\partial \delta_{\mathrm{i}}(\mvec{k}) }{\partial \delta_{\mathrm{q}}(\mvec{x})} = \sum_\mathrm{a} \mathcal{\bar{F}}_{\mathrm{i},\mathrm{a}}\delta^\mathrm{K}_{\mathrm{a},\mathrm{q}} \, , \nonumber
\end{eqnarray}

The final expression is:

\begin{eqnarray}
\frac{\partial \mathrm{ln}(\pi(\delta))}{\partial \delta_\mathrm{q}} = \sum_{\mathrm{i}} \frac{\partial \mathrm{ln}(\pi(\delta))}{\partial \rho_{\mathrm{i}}} \times \nonumber \\ \times \, \bar{n}_{\mathrm{gal}} \left \{ b_1 + [\sum_\mathrm{b}\mathcal{F}_{\mathrm{p},\mathrm{b}}\delta^\mathrm{K}_{\mathrm{b},\mathrm{j}}][\Delta b(\mvec{k},\fnl) \delta^\mathrm{K}_{\mathrm{i},\mathrm{j}}][\sum_\mathrm{a} \mathcal{\bar{F}}_{\mathrm{i},\mathrm{a}}\delta^\mathrm{K}_{\mathrm{a},\mathrm{q}}] \right \}
\end{eqnarray}

\label{lastpage}

\end{document}